\documentclass[apj]{emulateapj}
\usepackage{graphicx}

\def\gtorder{\mathrel{\raise.3ex\hbox{$>$}\mkern-14mu
             \lower0.6ex\hbox{$\sim$}}}
\def\ltorder{\mathrel{\raise.3ex\hbox{$<$}\mkern-14mu
             \lower0.6ex\hbox{$\sim$}}}

\def\mearth{{\rm\,M_\oplus}}

\shorttitle{Terrestrial Planet Formation With Migration}
\shortauthors{Mandell, Raymond \& Sigurdsson}

\begin{document}
\title{Formation of Earth-like Planets During and After Giant Planet
Migration} 
\author{Avi M. Mandell\altaffilmark{1,3,4}, Sean N. Raymond\altaffilmark{2} \& Steinn Sigurdsson\altaffilmark{1}}
\altaffiltext{1}{Pennsylvania State University, Department of Astronomy \& Astrophysics,
University Park, PA 16803, USA}
\altaffiltext{2}{NASA Postdoctoal Program Fellow, Center for Astrophysics and Space Astronomy, University of Colorado, Boulder, CO 80309-0389, USA}
\altaffiltext{3}{NASA Goddard Space Flight Center, Code 693, Greenbelt, MD 20771, USA}
\altaffiltext{4}{Corresponding Email:  mandell@astro.psu.edu}

\begin{abstract}

Close-in giant planets are thought to have formed in the cold outer regions of
planetary systems and migrated inward, passing through the orbital
parameter space occupied by the terrestrial planets in our own Solar System.
We present dynamical simulations of the effects of a migrating giant
planet on a disk of protoplanetary material and the subsequent evolution of the planetary system. We numerically investigate the dynamics of post-migration planetary systems over 200 million years using models with a single migrating giant planet, one migrating and one non-migrating giant planet, and excluding the effects of a gas disk. Material that is shepherded in front of the migrating
giant planet by moving mean motion resonances accretes into "hot Earths", but survival of these bodies is strongly dependent on dynamical damping.
Furthermore, a significant amount of material scattered outward by the giant
planet survives in highly excited orbits; the orbits of these scattered bodies are 
then damped by gas drag and dynamical friction over the remaining accretion
time.  In all simulations Earth-mass planets accrete on approximately 100 Myr timescales, often with orbits in the Habitable Zone. These planets range in mass and water content, with both quantities increasing with the presence of a gas disk and decreasing with the presence of an outer giant planet. We use scaling arguments and previous results to derive a simple recipe that constrains which giant planet systems are able to form and harbor Earth-like planets in the Habitable Zone, demonstrating that roughly one third of the known planetary systems are potentially habitable.
\end{abstract}


\section{Introduction}
\label{sec:intro}

More than 200 giant planets are known to orbit main-sequence stars
(\citet{but06}; see \citet{sch06} for recent results) and all but 5 have
semi-major axes within the orbit of Jupiter; more than half of known planets
reside within 1 AU of their parent star.  There are also a surprising number
of planets at very small semi-major axes: 22\% of currently known extrasolar planets have orbital radii less than 0.1 AU, and 16\% are located within 0.05 AU of the
central star.  Limitations on the current observational techniques do not
allow a complete sample of planets around solar-type stars beyond
approximately 3 AU (see, e.g., \citet{tab02}), but it is clear that
there is a significant population of planetary systems with giant planets at
small orbital radii. The occurrence of close-in giant planets is surprising
because theoretical models predict that the formation of giant planets in the
hot inner regions of a protoplanetary disk would be difficult under realistic
conditions \citep{bod00}. To explain the observations of massive planets at
very small distances from their parent stars, a mechanism is necessary to move
them from where they formed to where they currently reside; this process is
commonly known as planetary migration \citep{lin96}.  With the inclusion of
migration of a giant planet to small orbital radii, theories on the evolution
of solid bodies in the inner disk developed for our own planetary system must
be re-examined.

\subsection{Giant Planet Formation and Migration}
\label{sec:gpform}  
Favored theories of giant planet formation center around two main paradigms,
commonly called the core accretion model and the gravitational instability
model.  The bottom-up core accretion model \citep{pol96,ali05,hub05} requires
the accretion of planetesimals into a solid core of $\sim$ 5-10 $\mearth$,
massive enough to initiate runaway gravitational infall of a large gaseous
envelope.  The formation of this massive core requires a high surface density
of solids, which is difficult to achieve in the hot inner disk. A jump in the
surface density is believed to occur just past the ``snow line'', where the
disk temperature drops below the freezing point of water (at $\sim$ 170 K in
protoplanetary disks \citep{hay81}) and ice is available as a building block.
Isolation masses also increase with orbital distance $r$ for surface density
profiles flatter than $r^{-2}$ \citep{lis87}, also favoring formation at
larger orbital distances.  The top-down gravitational instability model
\citep{bos98,bos00,may02,dur05} assumes a gaseous protoplanetary disk massive
and cold enough to become gravitationally unstable to collapse, resulting in a
massive gaseous planet with little or no solid core.  For realistic disk
masses, this process only becomes efficient at orbital radii greater than
$\sim$10 AU \citep{may02}. Both are believed to be viable under certain
physical conditions, but it is still unclear how each mechanism (or
combination of mechanisms) functions in realistic circumstellar environments;
even calculations of the core masses for the giant planets in our own system
have large uncertainties \citep{gui99,sau04}.  Regardless of which mechanism
forms planets in extrasolar planetary systems, it is clear that giant planets
detected as small orbital radii were unlikely to form at their present
locations.

A variety of potential migration mechanisms have been proposed to operate on
massive planets, including planet-planet scattering \citep{far80,wei96,ras96},
planetesimal scattering \citep{fer84,mur98}, and a several types of gas-planet
interactions (see \citet{pap06} for a review).  The most simple and robust
hypothesis is known as Type II migration.  Type II migration
\citep{pap84,lin96} works on planetary bodies that are massive enough to open
an azimuthal and vertical gap in the gas disk, locking the planet into common
orbital evolution with the gas disk.  Viscosity in the gaseous disk, thought
to be due primarily to the magneto-rotational instability (MRI, \citet{gam96}),
results in orbital decay and infall of the gas onto the central star (as
evidenced by accretion and stellar activity in young stars
\citep{muz03,cal04,eis05}).  Thus, the giant planet loses angular momentum and
migrates inwards coupled to the gas disk.  Simulations suggest Type II
migration timescales range between $10^5$ to a few times $10^5$ years,
depending on the disk and planet conditions \citep{nel04,dan03}.  Planets
halted at very small radii most likely ceased migration due to gas evacuation
around the central star or planetary mass loss onto the central star, creating
the "hot Jupiter" population (see \citet{pap06} for further explanation).
However, for orbital radii beyond approximately 0.1 AU these stopping
mechanisms would not be effective, and other explanations are required for
extrasolar planets detected in orbits between 0.1 and a few AU. The
intermediate stopping distances may be a result of dissipation of the gas disk
during migration \citep{tri98} or a "dead zone" where the MRI is ineffective
\citep{gam96}, but these mechanisms rely on extreme fine-tuning of parameters
or questionable physical conditions to explain the distribution of orbital
radii for extrasolar planets.

\subsection{Terrestrial Planet Formation}
\label{sec:tpform}
Standard theories of the evolution of a planetary system suggest that a
circumstellar disk will proceed through various accretionary stages
culminating in the final architecture of a stable planetary configuration.
Coagulation of the inceptive dust particles occurs through collisional
sticking until the solid bodies become massive enough to decouple from the
surrounding gas disk and settle to the disk midplane, resulting in meter-sized
objects on timescales of $10^4$ years \citep{lis93,bec00}. In this size range
orbital decay due to gas drag is very rapid \citep{wei77} and accretion must
happen very quickly to avoid infall onto the central star and overly extensive
mixing. The accretion rate can be enhanced by differential migration rates
\citep{wei77}, gravitational instability \citep{gol73,you02}, and/or
concentration of meter-sized bodies in spiral density waves generated in the
gaseous disk \citep{ric04}. Once the largest bodies reach $\sim$1 km in size,
their gravitational cross-section becomes larger than their geometric
cross-section and they begin the phase known as "runaway growth".  During this
phase the largest bodies grow faster than the smaller bodies due to
gravitational focusing and damping due to dynamical friction, resulting in a widening mass dispersion \citep{gre78,wet89,wei97}.  "Runaway growth"
transitions into "oligarchic growth" when the velocity dispersion of
planetesimals becomes comparable to the escape speed of the largest embryos
\citep{kok98}, stalling accretion for the largest bodies and decreasing the
embryo mass dispersion.  Oligarchic growth ends after approximately $10^{\bf
6-7}$ years when the oligarchs have depleted their "feeding zones"
sufficiently such that dynamical friction is no longer effective, and embryos
begin to scatter each other \citep{kok98,kok00}.  The final "chaotic phase" of
planet growth proceeds through scattering and collisions between the large
protoplanets and final clearing of the remaining planetesimals to produce a
stable planetary system after more than $10^8$ years
\citep{wet96,cha01,ray06c,ken06}.

Though the evolutionary state of the protoplanetary disk at the inception of
giant planet migration is uncertain, there are several reasons to believe that
relatively large objects would exist in the inner disk by the time migration
begins.  Recent giant planet formation models assuming core accretion as the
dominant mechanism give full formation timescales ranging from 5 Myr to less than 1 Myr \citep{ric03,ali05,hub05}, corresponding to the oligarchic growth stage in
standard terrestrial planet models.  Observations of gas and dust in
primordial disks give disk lifetimes of $\sim10^7$ years for gas
\citep{bri01,hai01}and less than 3 Myr for dust \citep{sil06}, suggesting that
solid bodies form long before the gas disk dissipates.  For our own Solar
System, radioactive dating of meteorites give equivalent ages for both the
earliest chondrule formation and differentiation in asteroidal bodies in our
own Solar System, supporting rapid evolution of large solid bodies
\citep{kle02}.  Timescales for giant planet formation by gas instability may
be as short as $10^3$ years, but in the massive disks required for gas
instability to function the migration mechanisms may be halted until much of
the gas has dissipated \citep{bos05}.  Likewise, if massive disks are indeed
required for giant planets to form quickly enough for migration to occur,
formation timescales in other parts of the disk as would most likely be
shortened as well.

One of the most interesting problems in the formation of planetary systems is
the distribution of volatiles compared with refractory materials in different
planetary bodies. Naturally, the final composition of planets formed through
the scenario described above depends on both the initial distribution of
material in the protoplanetary nebula and the subsequent dynamical evolution
of the system. The initial composition of solids in a circumstellar disk is
expected to follow a basic condensation sequence \citep{gro72}, but the
effects of radial transport of grains and mm to meter-sized objects through
turbulence or radial drift have the potential to dramatically re-arrange this
early compositional gradient \citep{ste88,ste97,cie06}.  However, evidence
from our own system suggests that late-stage protoplanetary material may
maintain the gross characteristics of the basic condensation sequence: analysis of
chondrites in meteorites from parent bodies in different regions of the inner
solar system suggest a constant increase from almost no water at 1 AU to
almost 10\% water at 2.5 AU \citep{abe00}, with cometary bodies thought
to contain at least half of their mass in water ice .  The dry nature of asteroidal
material in the vicinity of Earth further suggests that continued delivery of
water-rich material must have been necessary to produce the current water
inventory, either through comet impacts \citep{owe95} or through accretion of
disrupted asteroidal material \citep{mor00}.  The importance of radial
transport of material in defining the final composition of terrestrial planets
has been confirmed by N-body simulations, which suggest that late-stage
accretion of water-rich material is largely influenced by the characteristics
of giant planets in these systems, notably their orbital eccentricities
\citep{cha02,ray04,ray06}.

\subsection{Previous Work}
\label{sec:prev}
With the uncertainty inherent in the cornucopia of forces that may be
functioning on objects in various different size regimes in circumstellar
disks, untangling the complex dynamics during the formation and evolution of
solid bodies in the inner regions of a protoplanetary disk becomes difficult.
Several studies concerning the habitability of planetary systems in a variety
of conditions made the simple assumption that systems with close-in giant
planets could not be habitable \citep{lin01,gon01}.  Given the 'hot
Jupiter'-stellar metallicity correlation (e.g., \citet{fis05}), this placed
limits on the galactic locales likely to harbor Earth-like planets (the so-called 'Galactic Habitable Zone'; \citet{lin04}).  More recent studies
have concentrated on exploring the dynamical effects of giant planet migration
on a simplified disk, with varying results. \citet{arm03}, assuming
planetesimals in the inner disk would be destroyed by a migrating giant
planet, analyzed the evolution of the post-migration disk surface density with
a simple dust coagulation formulation. They found that if migration begins
after significant embryo formation has occurred, post-migration formation
would be unlikely due to dust depletion and insufficient remaining material.
In \citet{man03} we examined the assumption of planetesimal destruction during
migration by analyzing the probability of terrestrial-mass objects surviving
the inward migration of a giant planet.  We used dynamical modeling to
investigate the survival rate and the eccentricity and semi-major axis
distribution of objects using different initial orbital radii and different
giant planet migration rates, and concluded that up to 50\% of objects could
remain in the system, albeit with a large eccentricity distribution.
\citet{edg04} also examined the dynamical impact of the migration of a planet
of various sizes on a planetesimal disk, concluding that eccentricities could
rise as high as 0.4 but would be damped down quickly in the presence of gas.

Several studies have recently begun to investigate the evolution of a more
realistic protoplanetary disk, both during migration and in a post-migration
configuration. \citet{fog05} and \citet{fog06} analyzed the survival rate of inner-disk objects during migration in a range of size regimes approximating runaway growth,
including the effects of gas drag and subsequently a more realistic evolving disk model. They concluded that between 50 and 90\% of
material would be retained, with the distribution interior and exterior to the
giant planet varying with disk maturity. Alternately, \citet{ray05} examined
the formation and composition of terrestrial planets in the presence of an
existing inner giant planet at varying stopping distances and without a gas
disk, concluding that a hot Jupiter would have little influence on terrestrial
planet formation outside its orbit. That study proposed that terrestrial
planet formation would be inhibited for orbits with periods within a factor of
3 - 4 of either an outer or inner giant planet.

In this study we present detailed analysis of numerical simulations in which
we explore the final stages of terrestrial planet formation after the
migration of a giant planet. Simulations begin with a two-phase protoplanetary
disk and a fully-formed giant planet in the outer disk, and follow the system
through the giant planet's migration and 200 Myr of additional evolution,
sufficient to examine the final characteristics of planets in the terrestrial
zone. These results bridge the gap between the short-term effects of migration
on the planetesimal disk demonstrated by \citet{fog05} and the long-term
evolution and final configurations of planetary systems with close-in giant planets
explored by \citet{ray05}. A subset of these simulations were first presented
in \citet{ray06b} (referred to here as Paper I), and in this paper we present
two additional simulation sets run with different initial conditions and a
detailed description and analysis of all three simulation sets.

In Section \ref{sec:mod} we describe the simulation details and initial
conditions; in Section \ref{sec:res} we describe the results for all the
simulations and compare the three different models; in Section \ref{sec:xsp}
we explore the potential ramifications for known extrasolar planetary
systems; and in Section \ref{sec:conc} we provide a summary of our results and
suggest future work.

\section{Model Parameters}
\label{sec:mod}

\subsection{Initial Conditions}
\label{sec:initcond}
We performed three different sets of simulations to examine the effects of the
most important parameters in the simulations: the number and location of giant
planets and the presence of a gaseous disk during and after migration.  The
primary set includes only one Jupiter-mass planet that migrates, and includes
the viscous damping effects of a gaseous disk.  The second set has both a
migrating Jupiter-mass planet and a Saturn-mass planet stationary at 9.5 AU,
with viscous damping included.  The final set includes the two giant planets
but does not incorporate gas drag. Our three sets of simulations are named JD
for 'Jupiter and (gas) Drag', JSD for 'Jupiter, Saturn and Drag', and JSN for
'Jupiter, Saturn, No drag'. Using different random number seeds, we generated
five disks of randomly distributed embryos and planetesimals for each
set. However, one run in each simulation set was unusable due to corruption
in the data storage or transfers between processors; therefore final
results are presented here for four simulations in each set, resulting in a
total of twelve simulation results. The first four simulations (JD) were
presented in Paper I; we incorporate them here as part of the larger set of
simulations.

The disk models used in this study have three major components: a solid disk
of small protoplanetary bodies, a gaseous disk, and one or two fully-formed
giant planets.  We start our simulations at the end of the oligarchic growth
phase of the protoplanetary disk, such that $\sim$1000-km planetary embryos
have formed throughout the disk but the mass in km-sized planetesimals is
comparable to the mass in embryos \citep{kok98, kok00}.  Additionally, we
incorporate the dissipative damping from a decaying gas disk, which disappears
after 10 Myr \citep{bri01,hai01}. Finally, we include either one or two giant
planets, assuming them to be fully-formed by this stage due to rapid formation
processes in the outer disk. This is consistent with recent results showing
that giant planets may form in $<1$ Myr via either core-accretion
\citep{ric03, ali05, hub05} or gravitational collapse \citep{bos01,may02,mayer04}. 

The radial surface density profile of the solid material in the disk follows
from the minimum-mass solar nebula model (MMSN; \citet{wei77,hay81}):

\begin{equation}
\Sigma(r) = \Sigma_1 \, f \, \left(\frac{r}{1 AU}\right)^{-3/2}
\end{equation}

where $\Sigma_1$ is the surface density at 1 AU in an MMSN disk
($\sim6\,g/cm^{2}$ and $1700\,g/cm^{2}$ for the solid and gaseous components,
respectively; \citet{hay81}) and $f$ is a scaling factor for the total
surface density.  Since the presence of close-in giant planets correlates with stellar
metallicity (e.g. \citet{law03}), we model a significantly more massive
disk than the MMSN, with $f \approx 2.2$.  Similarly, the nebular gas density
follows an exponential profile of the form

\begin{equation}
\rho(r,z)=\rho_0(r)exp\left\{-z^2/z_0(r)^2\right\}\,g/cm^3
\end{equation}

as described by \citet{tho03}, where $\rho_0$ is the midplane density taken
from the MMSN:

 \begin{equation}
 \rho_0^{min}(r)=1.4\times10^{-9}(r/1\:\textsc{AU})^{-11/4}\;g/cm^3
 \end{equation}

and $z_0$ is disk vertical scale height given by

\begin{equation}
z_0(r)=0.0472(r/1\:\textsc{AU})^{5/4}\;\textsc{AU}
\end{equation}

The gas density decreases linearly over 10 Myr, simulating the removal of the
gaseous nebula on observed timescales \citep{bri01,hai01}.

\begin{figure}
\plotone{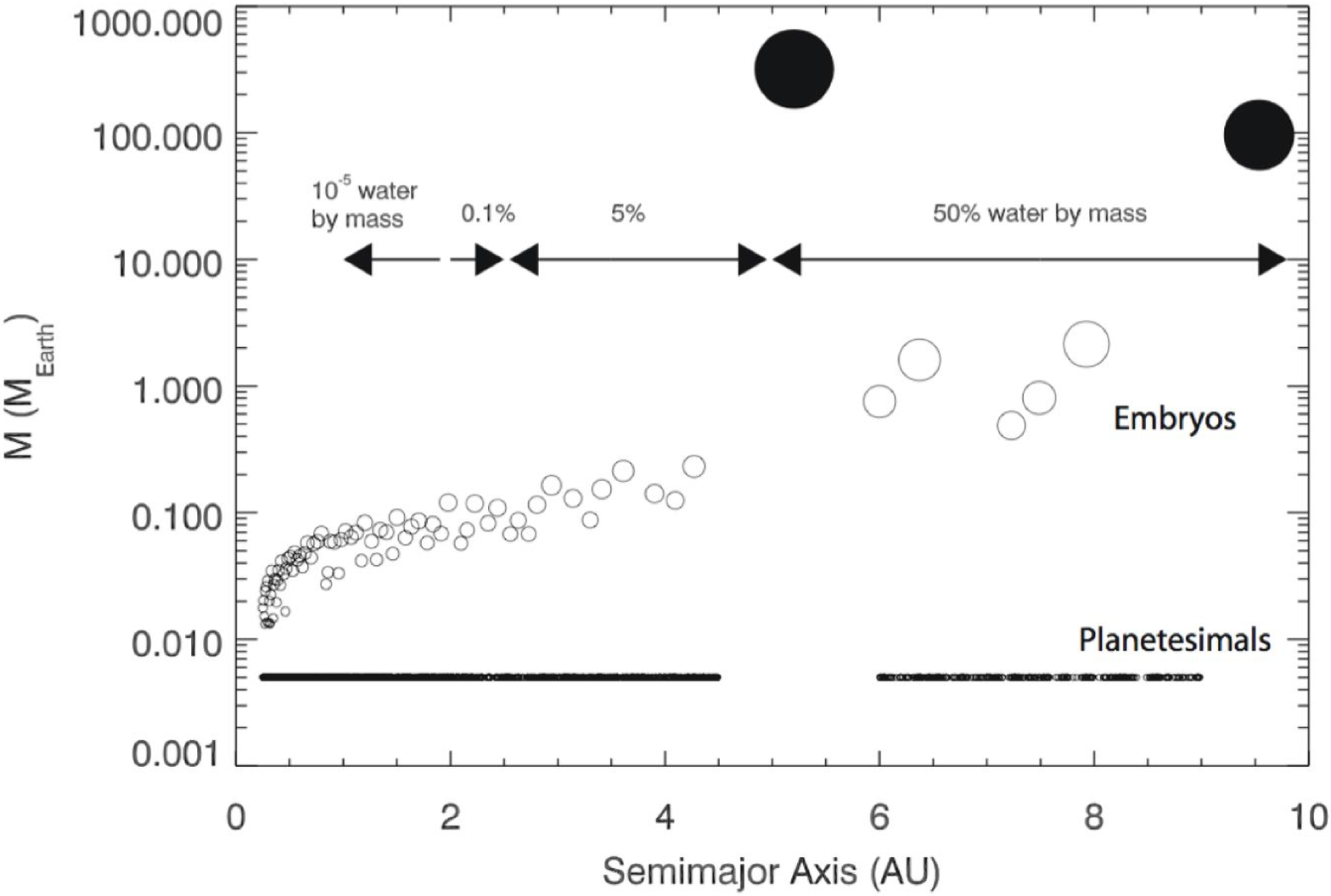}
\caption{Initial values for the mass and water mass fraction of protoplanetary disk used for each simulation. 
The solid disk is modeled after the \citet{hay81} MMSN, augmented by a factor of 2.2.  Embryos are spaced 
randomly by $\Delta$=5-10 mutual Hill radii, and embryo masses increase with radial distance $r$ as $M_{emb} \propto \Delta^{3/2}r^{3/4}$. Water and iron contents are based on measurements from Solar System bodies \citep{abe00,lod98}.}
\label{fig:init}
\end{figure}

Figure \ref{fig:init} illustrates our initial conditions for a single run.  The
initial solid disk is composed of $\sim$ 80 planetary embryos and 1200
planetesimals.  It extends from 0.25 to 4.5 AU (with a $\sim$ 2 Hill radii gap
on either side of the Jupiter-mass planet at 5.2 AU), and then from 6 AU to 9
AU.  The solid disk comprises 17 $\mearth$, of which 10 $\mearth$ is equally
distributed between planetesimals and embryos from 0.25 to 4.5 AU. Embryos
between 0.25 and 4.5 AU are spaced randomly by $\Delta$=5-10 mutual Hill
radii, and have masses between roughly 0.01 and 0.4 $\mearth$ with a mean
of 0.11 $\mearth$.  Embryo masses $M_{emb}$ increase with radial distance $r$ as $M_{emb} \propto \Delta^{3/2}
r^{3/4}$ \citep{ray04}.  In the inner region between 0.25 and 4.5 AU are 1000
planetesimals of 0.005 $\mearth$ each, distributed radially as $r^{-1/2}$,
i.e. as the annular mass in our surface density distribution. The surface
density distribution has a jump immediately past the snow line, assumed to lie
at 5 AU; therefore, isolation masses are larger in the region from 6 to 9 AU.
In this outer region we space embryos by $\Delta$=3-6 mutual Hill radii,
forming only 4 - 7 embryos between 0.2 and 1.1 $\mearth$, with a mean of
0.6 $\mearth$ and totaling 5.7 $\mearth$.  In
addition, 200 planetesimals comprising 1.3 $\mearth$ are placed in the region.
The higher embryo mass -  planetesimal mass ratio in the outer region
reflects the faster embryo formation time in this higher density region.
Indeed, massive embryos {\it must} have formed quickly in this scenario (if
giant planets form via core-accretion), because the giant planets are already
fully-formed.  Starting eccentricities are randomly selected up to 0.02 and
inclinations are set at 0.1$^{\circ}$, but these initial values are inconsequential
since the distributions are quickly perturbed.

Each embryo and planetesimal is assigned a composition based on its starting
location.  Water and iron contents are based on values from our Solar System,
where comets are thought to contain almost half their mass in water ice and 
asteroids beyond $\sim$ 2.5 AU contain significant quantities of water
(\citet{abe00}; see Section \ref{sec:tpform} and Fig. 2 from \citet{ray04}).  Inside 2 AU, embryos and
protoplanets are assumed to be dry, and beyond 5 AU they contain 50\% of their mass in water.
Between 5 AU and 2.5 AU they contain 5\% water by
mass, and from 2-2.5 AU they contain 0.1\% water by mass; this distribution
corresponds to starting mean water mass fraction of $\sim8\times 10^{-3}$ inside
5 AU (see Figure \ref{fig:init}).  Starting iron
contents are interpolated between the known values of the planets (values from
\citet{lod98}), including a dummy value of 0.4 in place of Mercury because of
its anomalously large iron content \citep{ray04, ray05};  the starting mean iron 
mass fraction is $\sim$0.32 inside 4.5 AU and $\sim$0.13 beyond 6 AU. 

Embryos and giant planets interact gravitationally with every body in the
system.  Planetesimals, however, are non self-interacting and are given an
effective mass to simulate a collection of much smaller particles
(e.g. \citet{tho03}).  In this way we can realistically include the effects of
1) gas drag on planetesimals and 2) dynamical friction of planetesimals on
embryos while using a reasonable particle number. Collisions are treated as
inelastic mergers conserving water and iron (for a discussion, see
\citet{ray04}).

\subsection{Simulation Details}
\label{sec:sim}
Numerical simulations were performed using a modified version of the
publicly-available hybrid symplectic integrator package MERCURY by Chambers \&
Migliorini \citep{cha97,cha99}.  We integrated each simulation for 200 Myr with
a 2-day timestep, which accurately integrates orbits with apocenter distances
larger than roughly 0.05 AU \citep{lev00}.  Energy is not properly conserved
for orbits inside 0.05 AU -- these bodies are typically given an artificial
energy ``kick'' that results in a close encounter and dynamical ejection.
This effect is important in certain instances and is discussed further in
Section \ref{sec:drag}.

To examine the effects of migration, we modified the integrator to accommodate
an artificial inward migration of a giant planet, as discussed in
\citet{man03}.  We use a simple drag force as described in \citet{chi02}
that produces a linear inward migration:

\begin{equation}
\mathbf{F}_{mig} = \frac{-M_P \, \mathbf{v}_P }{t_{mig}}
\end{equation}

where $M_P$ and $\mathbf{v}_P$ relate to the planet and $t_{mig}$ is the
migration time.  This method for simulating migration produces no artificial
changes in eccentricity and inclination for the migrating planet, and therefore
removes the potential for non-physical orbital excitation of the giant planet.
Migration rates as a function of time were modeled after recent simulations 
of Type II giant planet migration in the literature, which range from $10^5$ 
to $10^6$ years \citep{dan03,nel04} with an exponential tail to simulate 
potential braking processes (see Sect \ref{sec:intro}).

In addition, a basic fluid drag force was instituted to simulate the
dissipation of energy due to the surrounding gaseous disk. The gas drag takes
the basic form of Stokes drag:

\begin{equation}
\mathbf{a}_{drag} = -K v_{rel} \mathbf{v}_{rel}
\end{equation}

where $v_{rel}$ is the relative velocity of the object with respect to the
surrounding gas. The drag parameter is defined as

\begin{equation}
K = \frac{3 \rho_{gas} C_D}{8 \rho_m r_m}
\end{equation}

where $\rho_{gas}$ refers to the local gas density, $\rho_m$ and $r_m$ refer to the density and mass of the object and the drag coefficient $C_D$ is defined to be 1 \citep{ada76}.  The gas disk is assumed to revolve in a circular orbit at sub-Keplerian velocity due to internal pressure support, following the relation

\begin{equation}
v_{gas} = v_K (1 - \eta)
\end{equation}

with $\eta$ defined as

\begin{equation}
\eta = \frac{\pi}{16}(\alpha + \beta)\left(\frac{z_0(r)}{r}\right)^2
\end{equation}

where $\alpha$ and $\beta$ are the exponents of radial dependencies of density
and temperature respectively and $z_0$ is the radially-dependent vertical disk
scale height, defined below \citep{tho03}.  The solid bodies therefore
experience both a orbital damping effect and an inward migration, decreasing
with mass and increasing with gas density.

In these simulations we do not account for torques on sub-Saturnian sized bodies due to
density waves in the gas disk, which may lead to a reduced velocity dispersion (known as gravitational or tidal gas drag; \citet{war93}) and/or orbital decay (known as Type I migration; \citet{war97}) of embryo and planet-sized bodies.  We neglect these effects primarily because the role of disk torques on fully-embedded objects is still uncertain.  Type I migration rates from simulations range from $10^3$ to $10^6$ years \citep{tan02,dan03}, which would alternately result in either the loss of all solid bodies before the gas disk has disappeared, or very little change in orbital position for large bodies. It is even unclear whether migration or damping due to disk torques functions at all in realistic turbulent gas disks \citep{nel04,lau04}. \citet{mcn05} investigated the oligarchic stage of terrestrial planet formation for different Type I migration scenarios, demonstrating that faster migration rates serve to damp down excitation and widen embryo spacing for material within $\sim$ 2 AU.  This may prove to be important in determining the final mass and composition of terrestrial planets, but since the rate of migration of protoplanets compared with the migrating giant planet has a significant impact on the scattering energy and angular momentum transfer between the giant planets and the surrounding solid bodies, arbitrarily choosing a migration rate would strongly influence the results. We therefore chose not to include it in this work.  Future simulations will incorporate more detailed models of embryo migration taken from improved hydrodynamic simulations. 

\section{Results}
\label{sec:res}
The evolution of the protoplanetary disk can be separated into distinct
dynamical stages.  Migration (Stage 1) encompasses the migration period,
during which the giant planet moves through the interior regions of the disk.
Gas Dissipation (Stage 2) begins after giant-planet migration has ceased, and
continues until the remaining gas disk has fully dissipated. Clearing (Stage
3) includes the final stages of accretion and clearing in the absence of any
gaseous material.  In the following subsections we describe each stage, and
illustrate the differences between the results for the three different models
used in our simulations.  Snapshots during the different stages of the
evolution of each type of simulation are illustrated in Figures \ref{fig:jd}
(JD), \ref{fig:jsd} (JSD), and \ref{fig:jsn} (JSN).  Each plot follows the
evolution of all bodies in the simulation, and tracks their masses, orbital
parameters and compositions through time.  In addition, the mean and range 
of orbital and compositional properties for each model at the end of each stage
are listed in Tables \ref{tab:st1}, \ref{tab:st2}, and \ref{tab:st3}. 

\subsection{Stage 1: Migration}
\label{sec:stg1}
\begin{figure}
\plotone{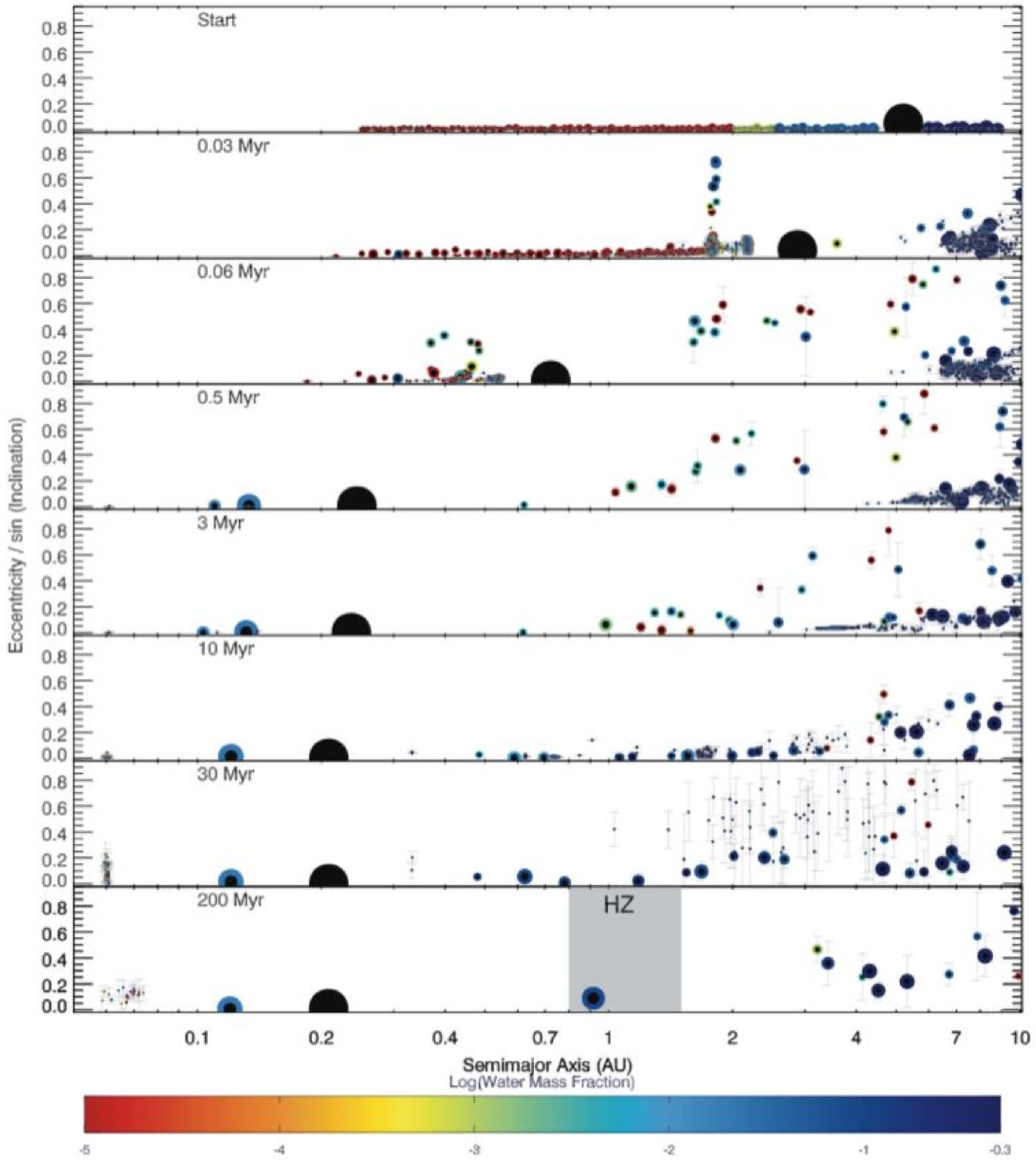}
\caption{Evolution of a sample simulation run using the JD model (a single migrating Jupiter-mass planet with gas drag included).  Radial distance is plotted on the x-axis, with eccentricity on the y-axis.  Relative inclination is indicated through error-bars on each point, and water content is indicated by the color (a
reference bar is located below the plot).  The size of each point is proportional to M$^{1/3}$, with the size of the inner black region representing the iron mass fraction.   The system forms a 'hot Earth' and a massive terrestrial planet in the Habitable Zone.  Water contents are high through radial mixing and migration of planetesimals. Reprinted from \citet{ray06b} (Paper I).}
\label{fig:jd}
\end{figure}

\begin{figure}
\plotone{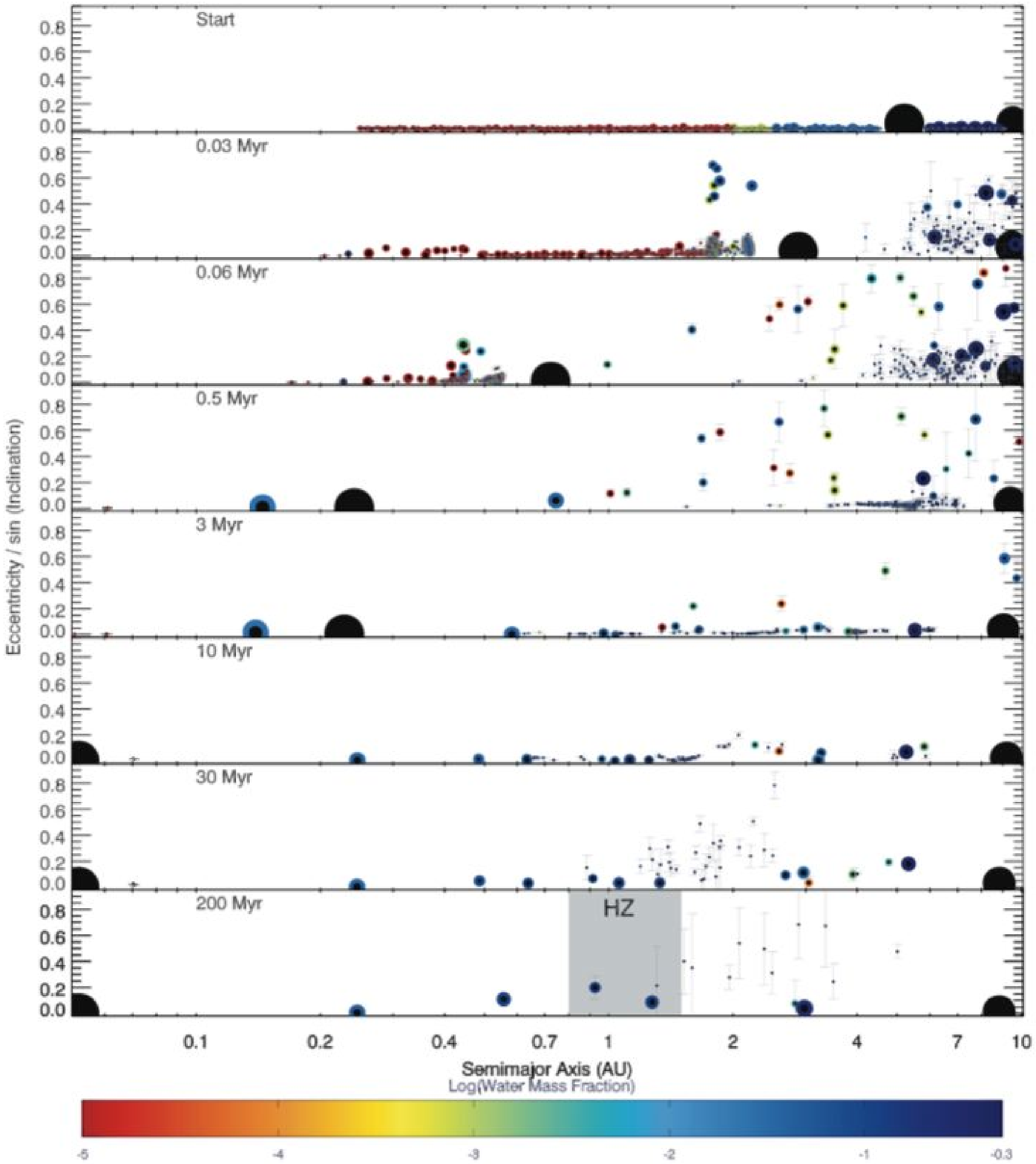}
\caption{Evolution of a sample simulation run using the JSD model (a migrating Jupiter-mass planet
with a second Saturn-mass outer planet and gas drag included).  The evolution on the inner system is similar
to the JD model, but the outer system is cleared rapidly and less water-rich material flows into the inner system.}
\label{fig:jsd}
\end{figure}

\begin{figure}
\plotone{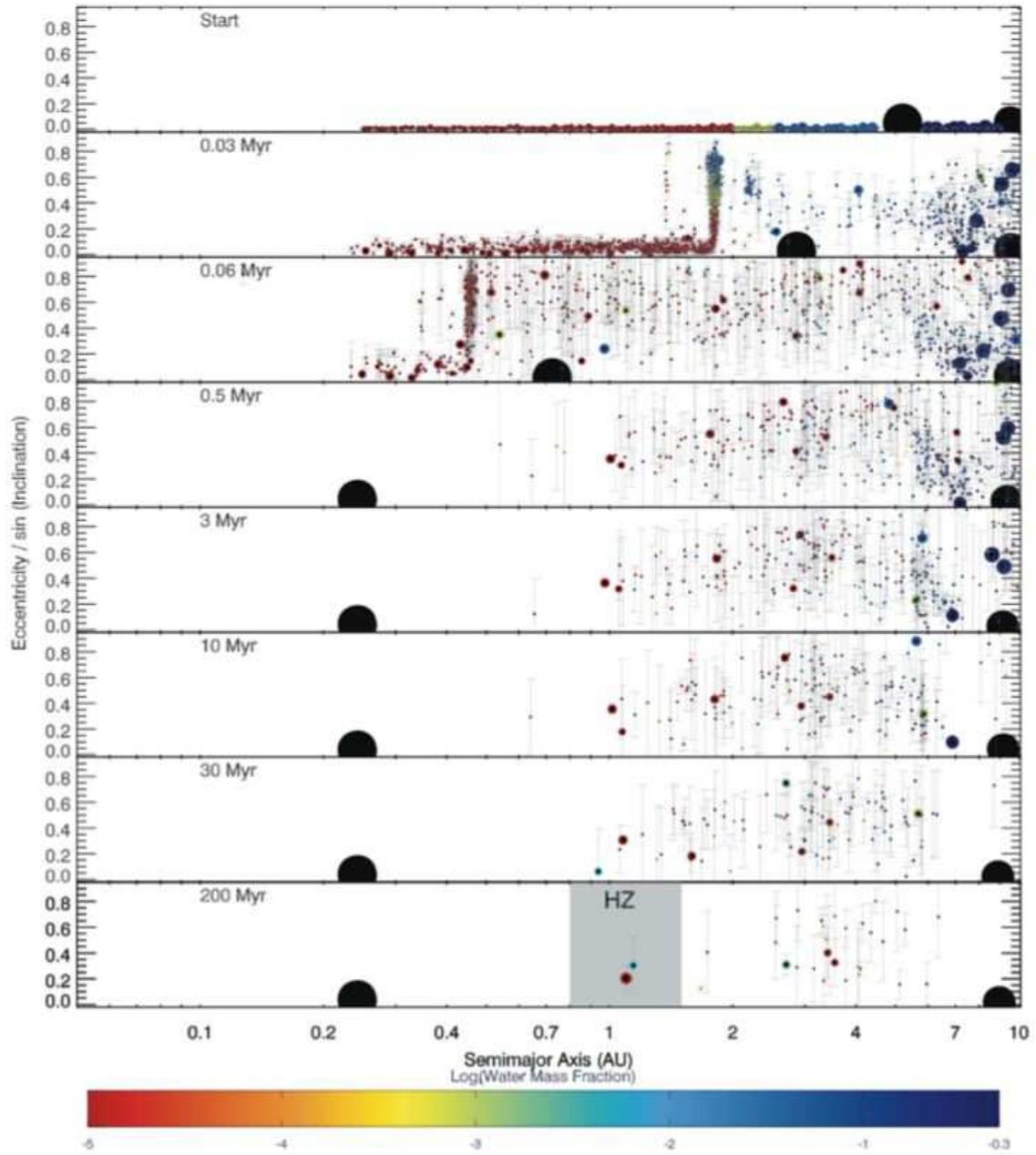}
\caption{Evolution of a sample simulation run using the JSN model (a migrating Jupiter-mass planet
with a second Saturn-mass outer planet and no gas drag).  The protoplanetary disk is highly excited, with
most the the mass being lost to ejection.  Over long time scales dynamical friction cools the disk, and low-mass
planets with lower water contents form.}
\label{fig:jsn}
\end{figure}

During Stage 1 the Jupiter-mass planet migrates inwards from 5.2 to 0.25
AU over $10^5$ years, and the inward progression of mean motion resonances
results in the shepherding and orbital excitation of the disk material
interior and exterior to the giant planet.  As the Jupiter-mass planet
continues through the inner disk, material is either scattered outward or
captured into resonant orbits and forced to migrate inward with the
Jupiter-mass planet.  This resonance capture, noted by \citet{tan99}, occurs
when a body close to a strong mean motion resonance (e.g., the 2:1 MMR) has
its orbit excited and its eccentricity increased.  Thus, the body's perihelion
distance has decreased and it has lost angular momentum.  An eccentric orbit
increases the relative velocity between the protoplanet and the
pressure-supported gas disk, thereby increasing the effects of gas drag. In
addition, an eccentric orbit encounters more nearby planetesimals and also
enhances dynamical friction.  These dissipative forces tend to reduce the
body's energy and decrease its semi-major axis and eccentricity.  Thus, the
protoplanet is moved inward of the resonance with the giant planet; as the
giant planet continues to migrate, the resonance can continue pushing it inward 
due to these angular momentum loss mechanisms.  Smaller bodies that feel gas 
drag more strongly can be pushed inward by higher-order resonances.  This is 
clearly seen in Fig. \ref{fig:jd}, where protoplanets are pushed inward by the 
2:1 and 3:2 resonances, and planetesimals by the 8:1 resonance. The decreasing 
spacing between resonances enhances the accretion and scattering rate as the 
giant planet moves inward, with planetesimal encounters peaking at approximately $7\times10^4$ years when the Jupiter-mass planet's semi-major axis reaches $\sim 0.4$ AU and the 2:1 resonance reaches the inner edge of the disk (see Figure \ref{fig:fate}).  At the end of the giant planet's inward 
migration, the remaining disk material is divided between bodies captured in 
very close low eccentricity orbits in interior resonances with the 
Jupiter-mass planet and bodies in high eccentricity orbits beyond 0.5 AU.

\begin{figure}
\plotone{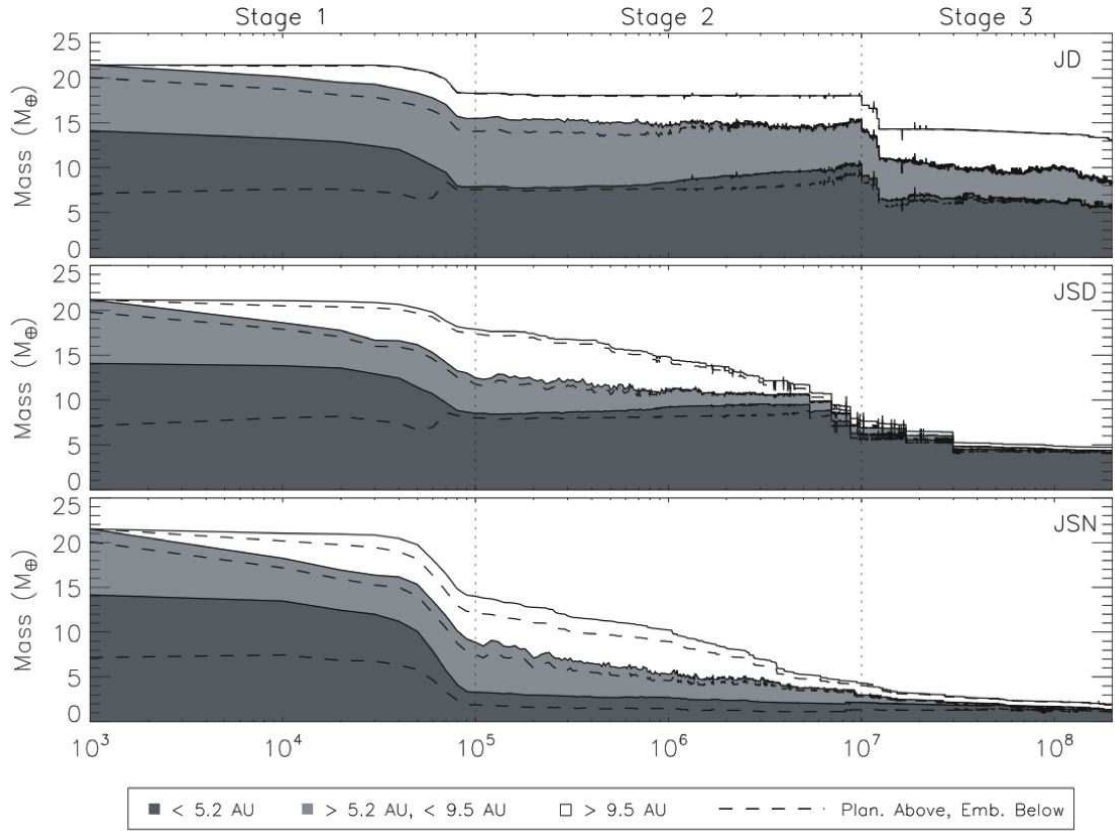}
\caption{The average total mass of the three models (in $\mearth$) versus time. The total mass is divided
into three zones: the inner disk (a $<$ 5.2AU), the intermediate disk (5.2AU $\le$ a $<$ 9.5AU), and the outer
disk (a $>$ 9.5AU).  Additionally, a dashed line is placed at the embryo / planetesimal mass boundary - below
the line is the mass from embryos, above it is the mass from planetesimals. Differences can be seen between the three models, specifically in the loss of mass
without the presence of gas and the rapid evolution of the outer disk with the presence of a second giant
planet.}
\label{fig:mass_3reg}
\end{figure}

\begin{figure}
\plotone{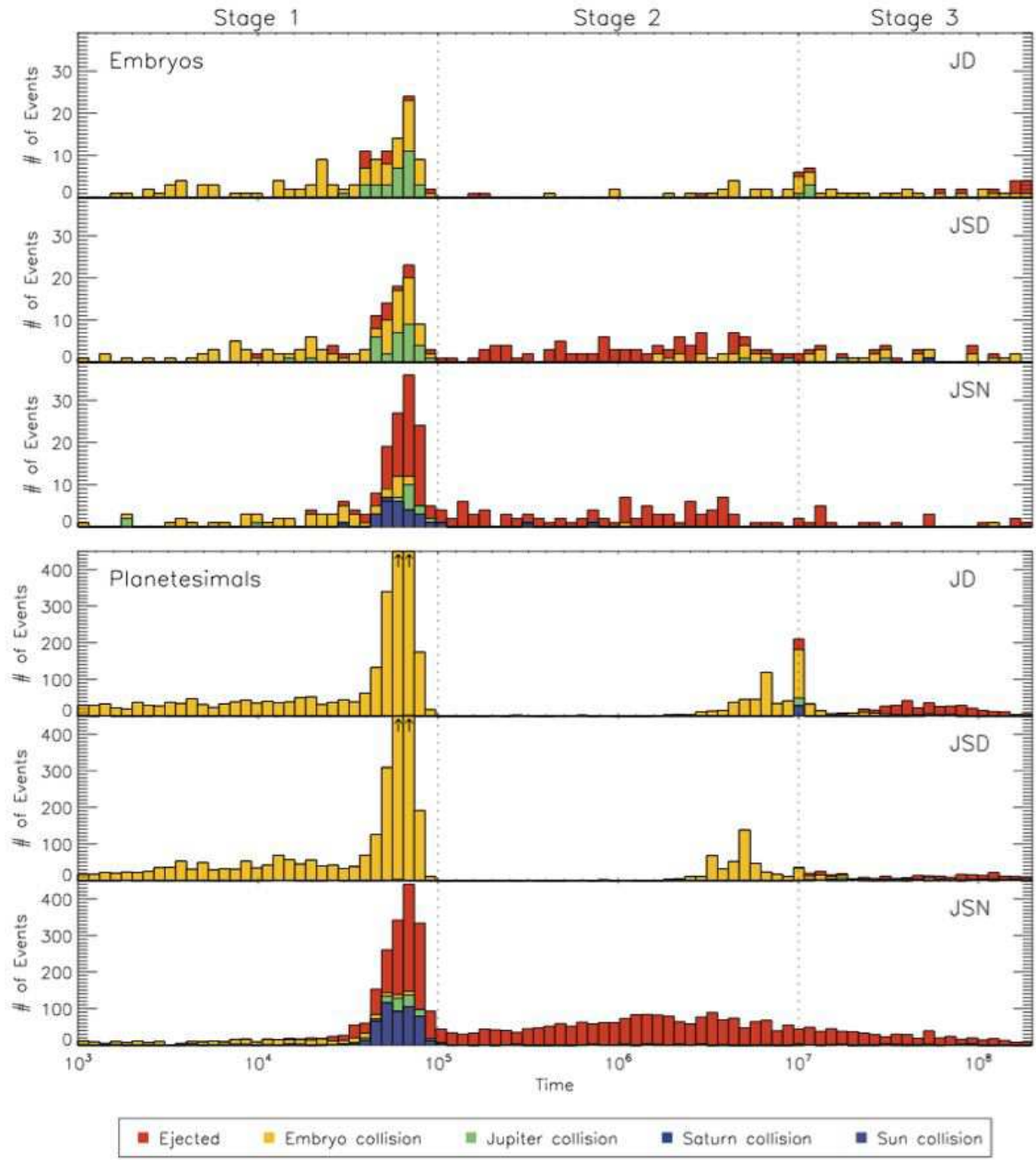}
\caption{Illustration of the loss processes for embryos and planetesimals over time for the three models.
Accretion happens primarily during migration, with additional accretion occurring as the gas disappears
at 10 Myr.  Simulations with gas drag show almost complete accretion onto embryos and Jupiter at early
times, then JSD simulations begin to eject bodies.  With gas, almost all mass is lost in ejections.}
\label{fig:fate}
\end{figure}

\begin{figure}
\plotone{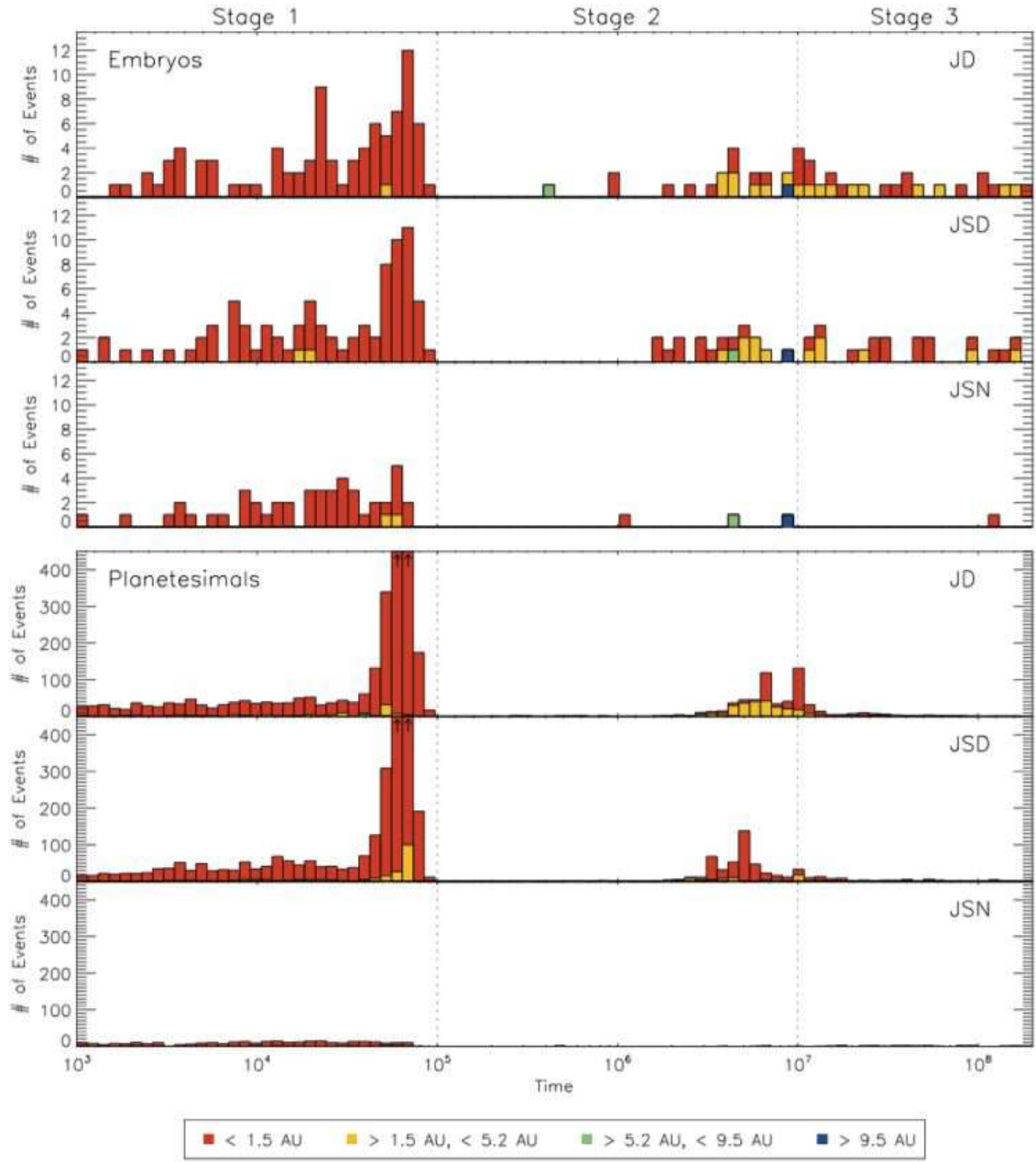}
\caption{Number of bodies accreted as a function of time, with the radial location of the impact denoted by
color, for the three different models.  Accretion occurs almost solely within 1.5 AU during migration, with additional accretion occurring at a range of distances as the gas disk dissipates.  Accretion is almost non-existent when
the gas is dense enough to damp down orbital excitation.}
\label{fig:accret}
\end{figure}

In the simulations including the presence of a gas disk (Models JD and JSD)
orbital excitation of planetesimals is almost completely damped in the inner
disk - by the end of Stage 1 the mean eccentricity for planetesimals within 9.5 AU is less than 0.1, compared with approximately 0.5 for embryos; mean inclination follows similar trends but with slightly higher values for bodies in the JSD simulations due to multiple scatterings (see Table \ref{tab:st1}).  The excitation of embryos is damped by a combination of relatively weak gas drag and dynamical friction from interactions with planetesimals. These effects are sufficiently strong to halt almost all
ejections, and the orbital radii of scattered bodies are limited to within 50
AU. Additionally, these forces result in strong resonant trapping by the
migrating planet, frequent collisions and a rapid increase in embryo
masses - Figure \ref{fig:accret} reveals that the large majority of both embryo-embryo collisions and embryo-planetesimal collisions occur during the migration period,
and Figure \ref{fig:mass_3reg} demonstrates the loss of planetesimal mass to embryo mass.  By the time migration has ended, the combination of rapid accretion
and scattering has cleared almost all planetesimals from within 5 AU, and the average embryo mass for this region has risen from 0.11 $\mearth$ in $\sim$65 objects to 0.45 $\mearth$ in $\sim$17 objects.  
Eccentricities for remaining embryos beyond 0.5 AU
are evenly distributed from 0.2 to 0.8, increasing with semi-major axis (see Figures \ref{fig:jd} and \ref{fig:jsd}); 
this relationship is characteristic of a scattered population of embryos as noted
by \citet{man03}.  Inclinations remain low due to damping, with average inclinations remaining below 5$^{\circ}$ (see Table \ref{tab:st1}).  Within 5 AU the mean composition of embryos does not change drastically since the bodies surviving in the inner disk are not augmented by any additional material over these short timescales, but the material is locally mixed.  The exception is the JSD model, where water-rich material is scattered inward by the outer Saturn-mass planet and the mean water mass fraction increases by a factor of 2.     

Beyond 5 AU the dynamical effects of the presence of a second giant planet
become important.  In simulations without a second planet, the only external
excitation for objects orbiting beyond the initial orbit of the migrating
Jupiter-mass planet comes from external resonance crossings and mutual
gravitational interactions.  Material native to this region remains relatively
undisturbed, with very little orbital excitation.  Bodies scattered into the region by the migrating giant have large eccentricities that are subsequently damped on Myr timescales. Maximum orbital eccentricity for these outer embryos increases up to $\sim 0.5$, and by the end of migration dynamical friction and damping due to gas drag reduce the maximum eccentricity to $\sim 0.2$ (see Figure \ref{fig:jd}).  In the JSD simulations, small bodies in the outer system are excited by both mean motion resonances and gravitational scattering.  This results in a significant scattered planetesimal population beyond 9.5 AU with a mean eccentricity of 0.3 and inclinations up to 30$^{\circ}$.  A small number of embryos are also scattered by both the migrating Jupiter-mass planet and the outer Saturn-mass planet, resulting in a semi-major axes as high as 50 AU.  

In addition to the presence of material remaining outside the orbit of the
close-in giant planet, all simulations including gas drag form one or more
terrestrial- mass bodies residing within the orbit of the Jupiter-mass planet (see 
Figure \ref{fig:hot}). These small bodies are the result of resonant shepherding by the giant planet as it migrates inward, and accretion at the resonance can build up bodies of several $\mearth$.  These ``Hot Earths'' are analogous to the 7.5 $\mearth$
planet found by radial velocity searches around the M star GJ 876 (Rivera et
al. 2005) and suggest that this phenomenon is common. Other computational
studies have demonstrated resonant shepherding as a mechanism to explain hot
Earths \citep{zho05,fog05, fog06}, though \citet{zho05} explain the planet around GJ
876 by invoking moving secular resonances during disk dissipation rather than
mean-motion resonances during migration.  However, in almost all of the simulations the inner 'hot Earth' becomes unstable as the gas disk disappears and either impacts the Jupiter-mass planet or is ejected.  Similarly, in simulations without any gas drag almost no accretion occurs interior to the migrating giant planet.  This sensitivity to damping forces was noted by \citet{fog06}, but may be mitigated by the presence of collisional debris not included in these simulations.  More detailed work is necessary to accurately model the long term stability of these bodies.

Simulations performed without the presence of a gas disk (Model
JSN) demonstrated very different encounters between bodies than Models JD and JSD throughout the disk.  Without the damping effects of gas drag on planetesimals, accretion
ceases after $\sim 40,000$ years as the Jupiter-mass planet enters the inner system (see Figure \ref{fig:accret}). Excitation from resonance crossings and direct scattering cause $\sim 40\%$ of the mass in small bodies to be lost from the system by the end of migration (see Figure \ref{fig:mass_3reg} and Figure \ref{fig:fate}), with the remainder of material distributed evenly in eccentricity and inclination phase space.  Approximately 65\% of the remaining mass still resides within 10 AU (compared to 75\% for simulations with gas drag), but the ratio of the total mass in embryos to the total mass in planetesimals in this region has only risen to only $\sim 2$, compared to $\sim 10$ for simulations including the effects of a gaseous disk.  The mean embryo mass is 0.18 $\mearth$ compared with 0.45 $\mearth$ for the simulations with gas drag, and the mean eccentricity and inclination for both embryos and planetesimals are both significantly higher than the JD and JSD simulations.

\subsection{Stage 2: Gas Dissipation}
\label{sec:stg2}
Once the Jupiter-mass planet has ceased migrating at 0.25 AU, the remaining
bodies in the disk continue to scatter and accrete each other during Stages 2
and 3, similar to the classical oligarchic and chaotic stages of terrestrial
planet formation.  At the start of Stage 2 the protoplanetary embryos in the disk are dynamically hot, with mean eccentricities increasing with orbital
radius.  However, during Stage 2 the presence of gas drag and dynamical friction serve to damp down eccentricities and inclinations and replenish the inner disk due to migration of planetesimals. During this stage the location and rate of accretion depends critically on the balance between the size of planetesimals and embryos and the density of the gas disk: faster migration leads to higher accretion rates at smaller orbital radii, while damping of dynamical excitation leads to reduced accretion and resonance capture.

After a hiatus of several million years due to clearing by the giant planet
migration and damping by the remaining gas disk, the accretion of planetesimals increases over the final duration of Stage 2 as a result of the refilling of feeding zones due to planetesimal orbital decay and increased orbital crossing due to decreased gas drag (see Figure \ref{fig:accret} for the accretion history of the combined data sets).  The accretion wavefront proceeds inward as the gas becomes less dense and orbital excitation increases.  This late-stage accretion is responsible for the high water mass fractions seen in the final configurations - embryos acquire most of their mass during Stage 1, but most of their water during Stage 2 (see Figure \ref{fig:mt}).In simulations with the JD model accretion continues until after the gas has disappeared, but the presence of a second massive planet in the JSD model leads to clearing between 5 - 15 AU and a decrease in accretion after $5\times10^6$ years. The outer planet also serves to clear out high-eccentricity embryos, leading to a much lower mean eccentricity and fewer embryos in the outer disk as compared with the JD model (see Table \ref{tab:st2}).  After 10 Myr the mass in planetesimals in both the JD and JSD models is only $\sim1 \mearth$, but the total mass in the JSD simulations is less than 30\% of the mass remaining in the JD simulations due to embryo ejection in the outer system (see Figure \ref{fig:mass_3reg}, Table \ref{tab:st2}). Eccentricities and inclinations in the inner system for both models decrease dramatically over the lifetime of the gas disk, with embryo eccentricities and inclinations decreasing by 20-50\% and planetesimals attaining a mean eccentricity below 0.1 and a mean inclination less than 2$^{\circ}$. 
   
\subsection{Stage 3:  Clearing}
\label{sec:stg3}
In the simulations with gas drag, Stage 3 begins as the final gas disk
dissipates and dynamical interactions begin to excite the remaining solid
bodies.  Accretion rates decrease steadily and cease by approximately 30 Myr, and scattering and ejection clear out most of the material in the inner disk over the remaining 170 Myr (see Figure \ref{fig:fate}).  At the end of the simulations only a few large protoplanets are left in the inner disk, but longer dynamical timescales in the outer disk result in longer clearing times and increased final eccentricities and inclinations for the JD simulations - the mean eccentricity rises to $\sim$0.3 and inclinations increase up to $\sim$50$^{\circ}$ (see Table \ref{tab:st3}). In contrast, the Saturn-mass planet begins to excite and scatter bodies near the end of Stage 2, and continues to eliminate highly excited bodies so that the mean eccentricities and inclinations for embryos at the end of 200 Myr are below 0.2 and 20$^{\circ}$ respectively.  The few remaining planetesimals are scattered randomly in eccentricity and inclination space.

In the JSN simulations, the system moves directly from Stage 1 to Stage 3.  Ejections of both embryos and planetesimals increase over the first 10 Myr, then begin to tail off (see Figure \ref{fig:fate}). Eccentricities maintain a flat distribution up to 0.7 due to a random distribution of scattering energies, while inclinations are spread out to 60$^{\circ}$.  More than 70\% of the total mass is lost by 5 Myr due to ejection and impacts with the central star, leaving only a small number of embryos in chaotic orbits and a scattered population of planetesimals (see Figure \ref{fig:jsn}).  By 10 Myr the total mass in embryos is only 3 times the mass in planetesimals, compared with a total embryo/planetesimal mass ratio greater than 10 for the simulations with gas drag included.  Highly excited bodies are cleared out of the inner system over the remainder of the simulation period and the mean eccentricity and inclination of the remaining embryos decrease to 0.25 and 17$^{\circ}$ respectively, but the planetesimals remain in highly inclined and eccentric orbits.

\subsection{Final Configurations}
\label{sec:final}
Figure \ref{fig:migall} shows the final configurations of all twelve of our
simulations, with the Solar System included for comparison.  The details of
the final planetary properties for bodies within 5 AU are listed in Tables \ref{tab:jd},
\ref{tab:jsd} and \ref{tab:jsn}.  As seen in previous simulations (e.g.,
\citet{cha01, ray04}), a diversity of outcomes exists within each group of
similar simulations.

Clear differences can be seen between the different sets of simulations, in
particular between simulations with (JD and JSD) and without (JSN) gas
drag. Both the mean planet mass and the mean water mass fraction in
simulations with drag is much higher than without drag.  These trends have two
root causes: 1) the survival of bodies scattered during migration depends
strongly on drag as a dissipative force, leaving more material in the system
when gas is present and promoting radial mixing, and 2) icy planetesimals from
beyond 5 AU can spiral in to the terrestrial zone over the lifetime of the gas
disk.  Simulations with drag also form large hot Earths due to
resonance capture by orbital damping (discussed in Section \ref{sec:stg1});
Figure \ref{fig:hot} demonstrates that all the hot Earths lie just inside
resonances with the hot Jupiter.  These hot Earths are rare in simulations
without drag - only simulation JSN-3 formed a dry, 0.22 $\mearth$ hot Earth at
0.05 AU.  We attribute the existence of this planet to damping from dynamical
friction during migration rather than gas drag, but scattering and ejection is
much more likely under these circumstances and the probability of a massive
body surviving is low.  Also, the final radius of the close-in giant planets' orbits are 
scattered in the gas drag simulations, due to the artificial migration caused by non-physical giant planet - planetesimal effects
(discussed in Section \ref{sec:drag}); in simulations without drag the giant planet
remains at 0.25 AU.

There also exist differences between simulations with and without an outer
giant planet.  The JD simulations contain extensive scattered disks beyond a few AU
with high inclinations and eccentricities and very long accretion timescales,
reminiscent of the Solar System's scattered disk of comets beyond Neptune
\citep{luu97, dun97}.  These disks are likely to produce vast amounts of cold
dust and may be detectable around other stars; late-stage accretion and
fragmentation due to this scenario could be responsible for the debris disks
seen around intermediate-age main-sequence stars such as $\beta$ Pictoris
\citep{gol06,bry06}.  Outer disk material is cleared out in simulations with
an outer giant planet, greatly reducing the amount of surviving outer disk
material, suggesting that dust mass production could correlate positively with
giant planet migration and negatively with giant planet survival in the outer
system.

Potentially habitable planets survive in almost all of the simulations.  In the eight
simulations with drag, seven planets larger than 0.2 $\mearth$ (including two in simulation JSD-4) formed in the Habitable Zone, the orbital region where 
the stellar flux is sufficient to maintain liquid water on the surface of a planet 
(assumed to lie between 0.8 and 1.5 AU for these simulations; \citet{kas93}).  Five planets with
masses from 0.13 to 0.4 $\mearth$ formed in the Habitable Zone in the four
simulations without drag.  In two separate cases, however, two planets remain
in the Habitable Zone on crossing orbits (sims. JSN-2 and JSN-5).  In each
case both of the planets have inclinations of $\sim20^{\circ}$ so the
timescale for the two planets to collide is long. Long-term dynamical studies
are necessary to confirm the stability of these planets on billion-year
timescales.  Planets as large as 1 $\mearth$ form within 5 AU in simulations
without gas drag, but it seems clear that additional material (either through
a more massive disk or an inward flux) is necessary to form similarly sized planets in the
Habitable Zone. 

\subsection{Drag-Induced Inward Migration}
\label{sec:drag}
In some cases, the inner gas giant and/or surviving embryos underwent
additional migration during the course of Stage 2. The cause of this
additional migration originates with the resonance capture of planetesimals
spiraling inward due to gas drag.  In most cases, the planetesimals become
trapped in outer resonances, then cascade down to just outside the 2:1 mean
motion resonance with the giant planet. This outer resonance capture was
investigated by \citet{tho05} with regard to larger objects, focusing on
differential migration due to differences in Type II (giant planet) and Type I
(terrestrial planet) migration rates.  In this case, the inward migration
comes from aerodynamic drag, and the capture process is aided by damping of eccentricities by the gas disk. The effect is strongest in the
inner disk where gas densities are high, and once the gas disk dissipates the
planetesimals are excited and either accreted onto the larger object or
ejected from the system.

\begin{figure}
\plotone{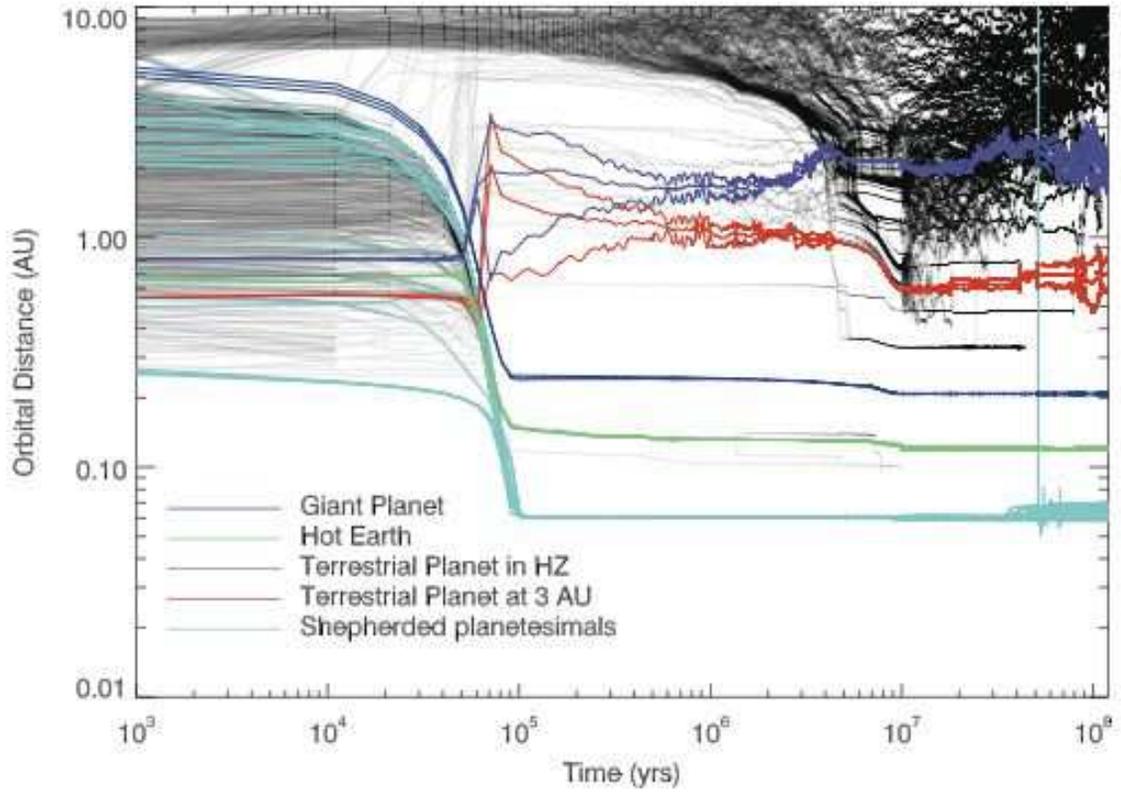}
\caption{Plot of orbital distance with time for planetesimals (black) in a sample JD simulation. Colored lines denote the evolution of the initial accretion seeds for important bodies in the simulation. As the Jupiter-mass planet moves inward, planetesimals and embryos are shepherded in resonances to become 'hot Earths'.  Other embryos are scattered outward to become outer terrestrial planets.  Multiple lines show the range in orbital distance due to eccentricity.}
\label{fig:clj}
\end{figure}

\begin{figure}
\plotone{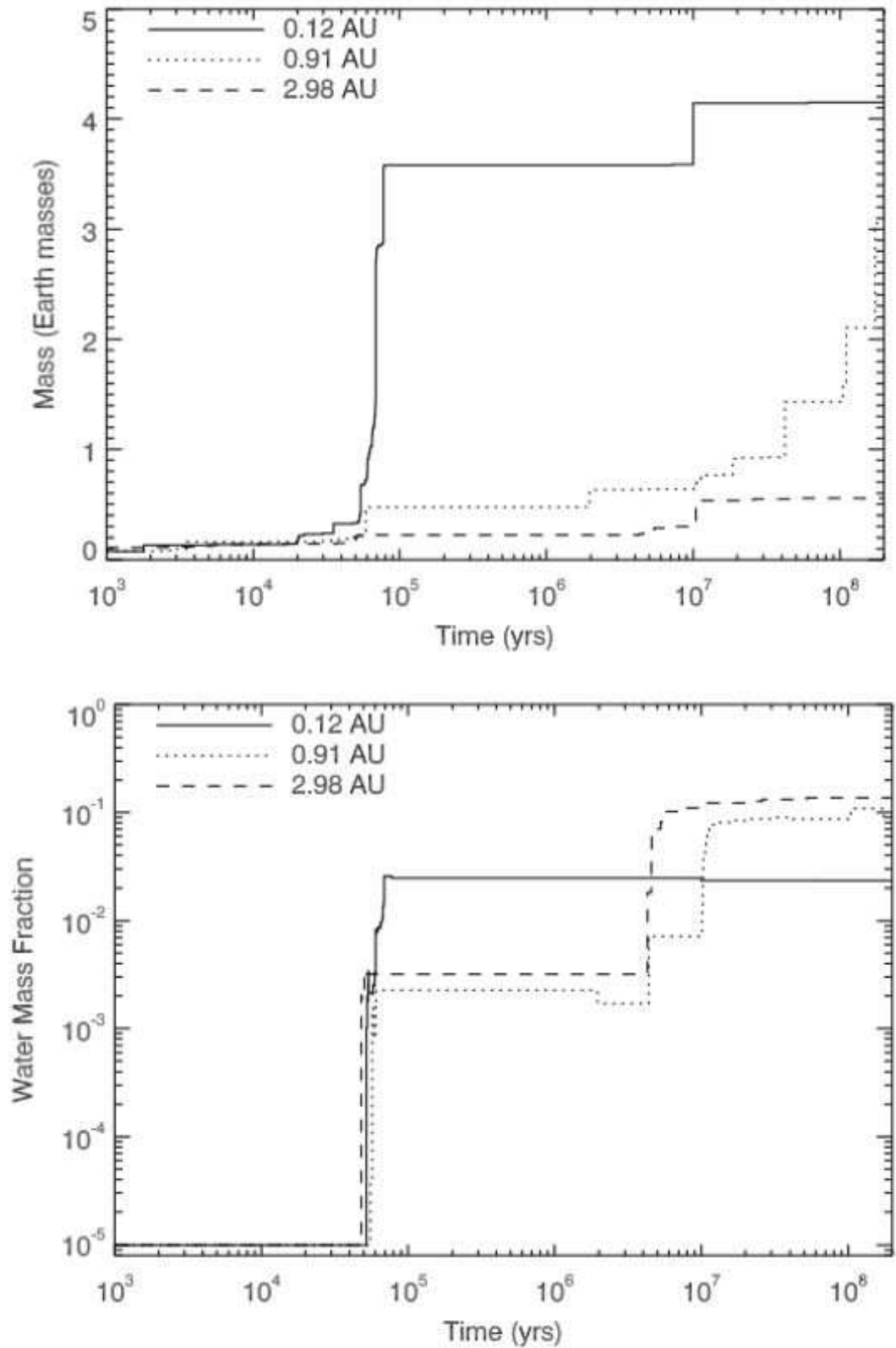}
\caption{Plot of mass versus time (top) and water mass fraction versus time (bottom) for three indicative bodies in a sample JD simulation.  The body that settles
at 0.12 AU (a 'hot Earth') accretes very rapidly near the end of migration, while the outer terrestrial planets 
accrete material gradually over longer timescales.  For the outer terrestrial planets, most of their water is accreted at two distinct times: just before scattering near the end of migration (Stage 1), and at the end of gas dissipation (Stage 2) when orbital excitation increases.}
\label{fig:mt}
\end{figure}

\begin{figure}
\plotone{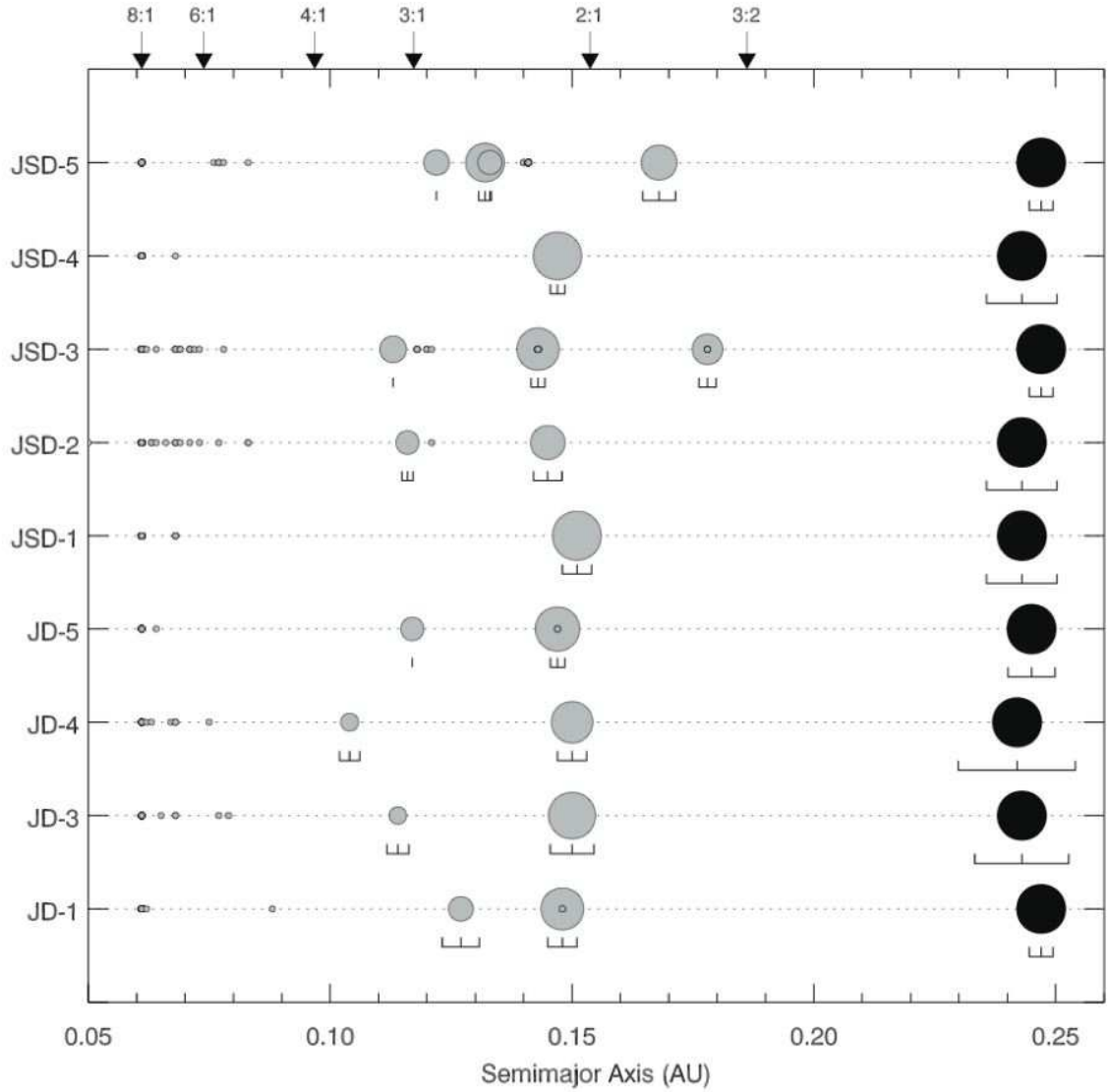}
\caption{Final configurations for 'hot Earths' in all simulations with gas drag (JD and JSD) before instability
sets in when the gas disk dissipates.  Resonance positions for a 'hot Jupiter' at 0.25 AU are marked.  The size of each body corresponds to its mass, with the orbital range (i.e. eccentricity) indicated below.  It is clear that most 'hot Earths' reside just inside resonances, due to the combined shepherding effect of the resonance and the orbital decay from gas drag.}
\label{fig:hot}
\end{figure}

\begin{figure}
\plotone{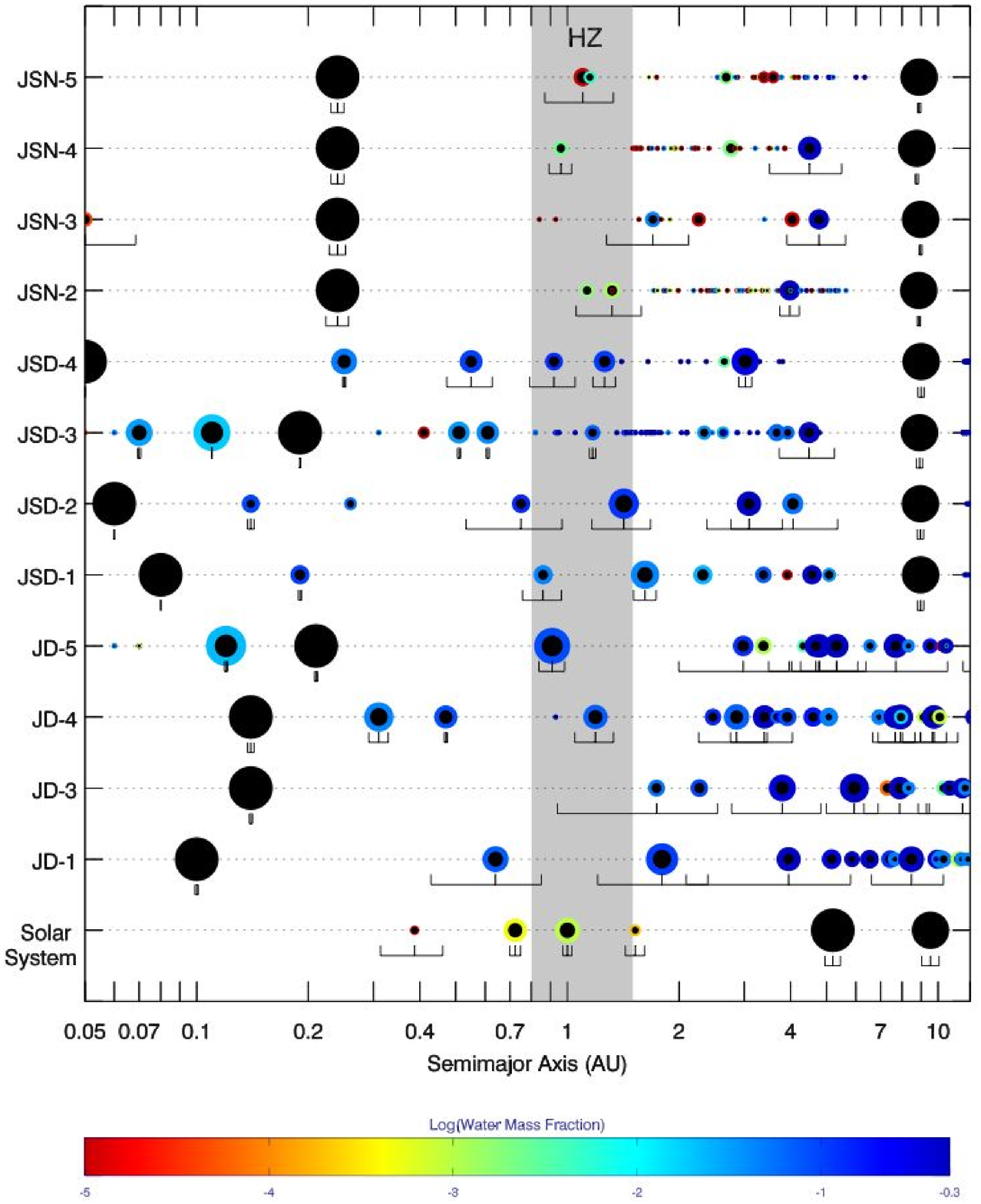}
\caption{Final configurations for all the simulations after 200 Myr of integration. The size of each body indicates mass and the color indicates
water content (with the Solar System plotted on the bottom for reference).  The orbital range (i.e. eccentricity) for each body is indicated by the horizontal error-bar.  It is clear that systems with long-lived gas disks will form more
massive, water-rich planets in or near the Habitable Zone, but that planets will form even in systems without gas.  Additionally, dense gas in the inner system will lead to 'hot Earths' near resonances with the 'hot Jupiter'.}
\label{fig:migall}
\end{figure}

The combination of angular momentum loss from the larger object to the
planetesimals (through orbital excitation from the strong resonance
interaction) and transfer of this angular momentum from the planetesimals to
the gas disk (in the form of orbital damping) leads to a net loss of angular
momentum from the solid bodies and subsequent orbital decay.  The amount of
inward migration varied from simulation to simulation, depending on the
orbital radius and number of planetesimals trapped in resonance.  The total
amount of inward movement of giant planets ranged from zero to almost 0.2 AU
-- in most cases it was about 0.1 AU. For terrestrial embryos at 1 AU the
inward migration was similar, but no additional inward motion was apparent for
objects beyond 1.5 AU (see Figure \ref{fig:clj} for examples).  The effect on the final orbits of the terrestrial bodies was usually overshadowed by orbital rearrangement due to late impacts and scattering events. The additional inward motion of the inner Jupiter-mass planet also caused the orbits of close-in terrestrial planets in some
simulations to become too small to be properly resolved with our 2 day
timestep (\cite{lev00}).  These orbits became unstable and usually collided
with the close-in giant planet.

Though this additional source of inward migration represents an interesting
avenue for future work, the effect cannot be properly analyzed in this study
due to our unusual treatment of the physical qualities of the planetesimals.
As noted in Section \ref{sec:sim}, the planetesimals have a dynamical mass of
a body with a size of $\sim$1000 km, but interact with the gas disk with a
more realistic cross section of a collection of 10 km bodies. Thus, the rate
of angular momentum transfer for a single planetesimal caught in resonance is
equivalent to that of 10$^6$ smaller bodies. Additionally, for computational
efficiency planetesimals do not self-interact. Thus, when trapped in the same
resonant orbit these small bodies are unable to coagulate into a larger body
that would lose angular momentum to the gas disk less efficiently. The angular
momentum loss and subsequent orbital decay over the duration of the gas disk
is therefore much greater than would be present in a realistic system.

\subsection{Water Delivery}
\label{sec:water}

Water-rich material is accumulated by the growing planets throughout their
accretion.  As shown in Paper 1, two primary mechanisms contribute: 1) radial
mixing induced by large eccentricities as a result of scattering interactions with
the migrating giant planet; and 2) accretion of in-spiraling icy material
because of gas drag.  The first mechanism delivers water mainly in the form of
hydrated asteroidal material and takes place through each stage, although the
vast majority of radial mixing occurs through giant planet scattering during
Stage 1, with additional embryo-planetesimal scattering occurring during Stage
3 after the gas disk has dissipated.  This mode of transport is similar to the
model of \citet{mor00}, except that it is substantially enhanced by the large
eccentricities induced by close encounters between the migrating giant planet
and the smaller bodies.  The second mechanism occurs mostly during Stage 2 as water-rich material filters through the inner system over long timescales.  In
our own system this mechanism most likely contributed little if any water to
the Earth because of Jupiter's presence: in-spiraling icy bodies would have
first encountered Jupiter and been scattered outward or accreted.  In the
simulations presented here, the very water-rich planetesimals originating past
5 AU move inward quickly, but damping of excitation diminishes collisions with
embryos and allows the small bodies to reach sub-AU radii.  Accretion of these
bodies tends to occur shortly after gas dissipation when damping ceases and
their eccentricities and inclinations are increased by perturbation from
nearby terrestrial bodies, resulting in collisional orbits.

As discussed in Paper 1 and below, the water contents of Habitable Zone
planets in the simulations with gas drag are very high, with typical values of
$\sim$ 10\% water by mass.  However, it is difficult to translate these into
likely values, since we do not take water depletion from impacts
\citep{gen05,can06} or hydrodynamic escape \citep{mat86} into account.
Previous simulations with a single, non-migrating outer giant planet form
planets in the Habitable Zone with $\sim 5\times 10^{-3}$ water by mass
(Raymond et al 2004, 2006a).  If we assume that previous simulations would
have resulted in water contents similar to the Earth's ($\sim 10^{-3}$ by
mass) if depletion were accounted for, then the present simulations should
still have about 20 times as much water as the Earth.  However, it is possible
that previous simulations instead formed planets which would have more water
than the Earth because the giant planets in those simulations were on circular
orbits rather than mildly eccentric orbits similar to the orbits of our own
giant planets \citep{cha02,ray04}.  Thus, our current planets may in fact
contain significantly more water and are likely to be covered in global oceans
several km deep and be veritable ``ocean planets'' (e.g.,\citet{leg04}).

The water contents of Habitable Zone planets in simulations with and without
gas drag differ by about two orders of magnitude.  The reasons for this
contrast are tied to the presence of nebular gas, via radial mixing of
surviving bodies and in-spiraling of icy planetesimals.  Thus, the water
contents of our final planetary systems may in fact be dependent on our
assumptions about the lifetime and depletion of the gas disk and the effective
mass of planetesimals.  For instance, if gas disk dissipation is the halting
mechanism of the giant planet migration (as assumed by, e.g., \citet{tri98}),
then perhaps only a Jupiter-mass or so of gas remains at the end of migration.
We expect that re-circularization of scattered material and inward migration of
small bodies would continue to take place to some extent in such a disk, such
that a significant fraction of material would survive to repopulate the
terrestrial zone.  Thus, the amount of material accreted from the outer system
by planets in the interior of the disk would be less.  The terrestrial planets
in such systems would probably represent an intermediate case between our
simulations with and without drag, with moderate masses and water contents.
As shown by the final configurations, we can expect that such systems would
form habitable planets similar to those formed here.

\section{Earth-like Planets in Known Planetary Systems}
\label{sec:xsp}
It has been shown that an Earth-like planet would be dynamically stable for
long timescales in the Habitable Zones of many of the known systems of giant
planets (e.g., \citet{men03, jon05}).  However, these studies only test the
stability of existing particles over long timescales; it is assumed that
initially-stable objects would form in-situ.  This is not the case in
moderately unstable environments; stability models predict the existence of an
Earth-mass planet at $\sim$ 3 AU in the our system's Asteroid Belt, where
such a planet would indeed be dynamically stable.  Clearly, when assessing 
the potential for a viable location for habitable planets it is critical to model the 
actual formation process.  

Our results can be applied to constrain which of the known systems of
extra-solar giant planets could have formed a terrestrial planet in the
Habitable Zone. By combining limits on formation of terrestrial planets 
in the presence of various giant planet scenarios, we can make a more rigorous 
assessment of the possibility for terrestrial planets in known planetary systems 
and begin to develop criteria for a habitable system. In this section we combine 
limits on the formation of terrestrial planets in the presence of various giant 
planet scenarios from this study and a previous study, and assess of the probability
of the formation of terrestrial planets in known planetary systems.

Examining the typical spacing between terrestrial and inner giant planets in our
simulations, we find that a giant planet must be interior to approximately 0.5 AU for a planet
at the lower limit of habitability ($\sim0.3\mearth$; \citet{wil97}) to form inside the outer boundary of the Habitable Zone (we have assumed the outer edge of the
Habitable Zone to be 1.5 AU, although this value is uncertain and depends on the
condensation properties of CO$_2$ clouds \citep{kas93, for97,mis00}).  The
spacing of planets that form exterior to the close-in giant varies a great
deal from simulation to simulation.  In dynamical terms, what is relevant is
the ratio of orbital periods of the innermost planet with significant mass
($>$0.3 $\mearth$) and the close-in giant planet.  This value ranges from 3.3
to 43 in our simulations, with a mean [median] of 12[9].  Although there
clearly exists a range of outcomes, we can define a rough limit on the orbit
of an inner giant planet that allows a terrestrial planet to form in the
Habitable Zone (the most optimistic case, i.e. the closest spacing, puts the
giant planet at about 0.7 AU).
 
We suggest that terrestrial planets of significant mass can form in the
Habitable Zone of a Sun-like star if no giant planets exist between 0.5 and
2.5 AU.  We derive these values using the reasoning above based on our current
results, in combination with the results of \citet{ray06}, who used hundreds
of low-resolution accretion simulation to constrain which outer giant planets
can form Habitable Zone planets (see Fig. 2 from that paper).  These values
are clearly eccentricity dependent -- the characteristic spacing between giant
and terrestrial planets increases quickly with giant planet eccentricity
(e.g., \citet{ray06}).  The highest eccentricity of the close-in giant planets
in our simulation is about 0.1, and this value is similar for most
simulations.  For our inner limit, we therefore assume that a system with a giant planet
inside 0.5 AU and with an eccentricity less than 0.1 can form a planet in the
Habitable Zone.

We apply these rough limits to the known extrasolar planets.  Our sample consists of
207 planets in 178 planetary systems (with 21 multiple planet systems)
discovered by Aug 1 2006.  We include planets discovered via the radial
velocity, micro-lensing, transit and direct imaging techniques (data from
\citet{but06} via exoplanets.org, and \citet{sch06} via exoplanet.eu). We
exclude planets more massive than 15 Jupiter masses ($M_J$; roughly the brown
dwarf limit), unless they are part of multiple systems.  Because the extrasolar planet
host stars have a range in masses, we must calibrate our limits.  To do so, we
use a mass-luminosity relation that is a fit to data of \citet{hil04}:

\begin{equation} 
y = 4.101 x^3 + 8.162 x^2 + 7.108 x + 0.065,
 \end{equation}
 
\noindent where $y$ = log$_{10}\left(\rm L_\star / \rm L_\odot\right)$ and $x$
= log$_{10}\left(\rm M_\star / \rm M_\odot\right)$ (John Scalo 2006, personal
communication).  We assume a Habitable Zone of 0.8-1.5 AU around a solar-mass
star, and assume that its inner and outer limits scale with the stellar flux,
i.e., as $L_\star^{1/2}$.  We then assume that the dynamical scaling between
the Habitable Zone and our giant planet limits, measured in terms of orbital
period ratios, is independent of stellar mass.  Table~\ref{tab:gplim} lists
our inner and outer giant planet limits for a range of stellar masses.

\begin{figure}
\plotone{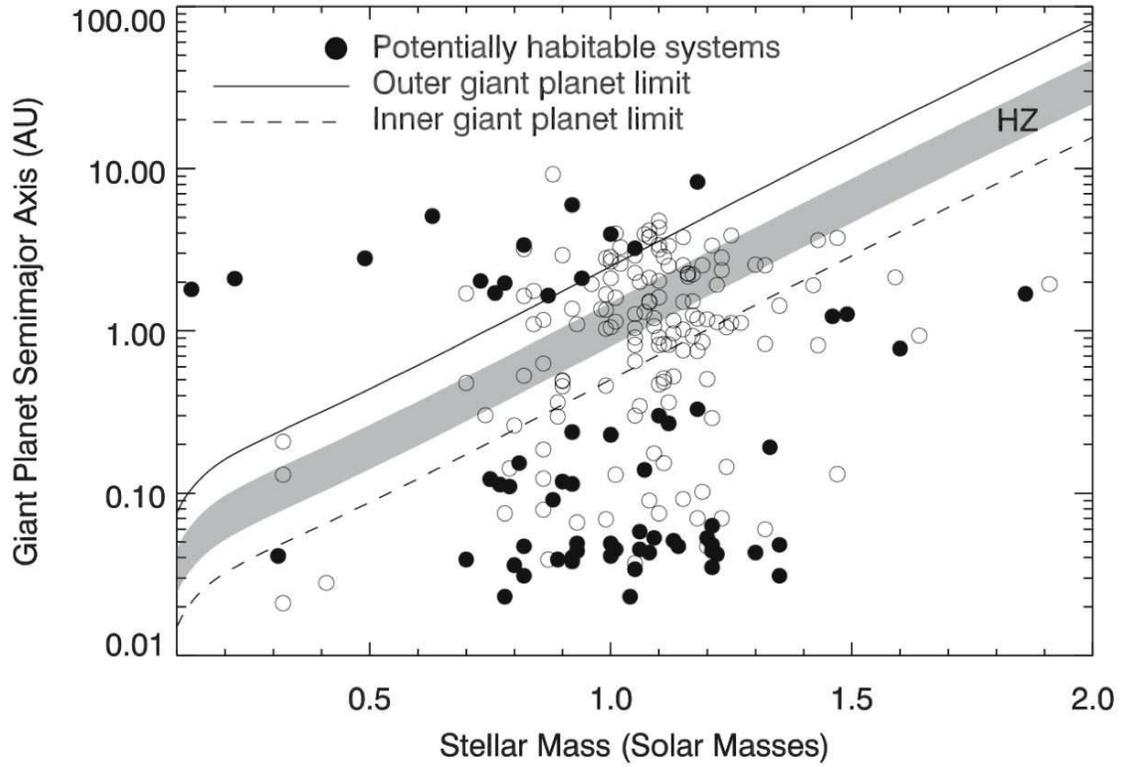}
\caption{The known extrasolar planets that are likely to have formed terrestrial
planets in the Habitable Zone.  The solid line shows the outer giant planet semi-major axis limit for habitable planet formation from \citet{ray06}, and the dashed line is the inner giant planet semi-major axis limit found from the current simulations.  Filled circles indicate the
known giant planets that satisfy our criteria (eccentricity less than 0.1 and semi-major axis within the derived limits) and may therefore harbor a terrestrial
planet in the Habitable Zone; open circles are unlikely candidates for
habitability.  Note that not all planets above or below our limits qualify,
due to the eccentricity limit.  The Habitable Zone is shaded. Modified from \citet{ray06b} (Paper I).}
\label{fig:xsp}
\end{figure}

Figure~\ref{fig:xsp} shows the known extrasolar planets that meet our criteria and are
good candidates for having a terrestrial planet in the Habitable Zone.
Potentially habitable systems are also listed in Table \ref{tab:xsp}, sorted
by host star mass.  Of the 178 systems in our sample, 65 (37\%) may have
formed a terrestrial planet of $0.3\mearth$ or more in the Habitable Zone and
so are considered to be potentially habitable systems.  Seventeen systems
satisfy our outer giant planet limit; these are Solar System-like in that they
have a giant planet exterior to the Habitable Zone.  In contrast, fifty
systems are decidedly different than the Solar system, with the potential for
an Earth-like planet coexisting with a close-in giant planet.  One system, 55
Cancri, falls into both categories, with three close-in giants and a distant
outer giant. Nine of the multiple planet systems contained a planet that
satisfied our limits but another planet that did not -- 55 Cnc was the only
multiple system for which all planets met our limits.
If we consider the ensemble of planets as a whole, then 75 out of 207 planets
(36 \%) met our criteria.  In the case of planets at very large orbital
distances, such as the ones detected by direct imaging (e.g., \citet{cha05}), the giant planet is so far away that it will be completely dynamically detached from the terrestrial region.

Our estimate that about one third of the known giant planet systems can harbor
potentially habitable planets diverges from previous estimates.  \citet{jon05}
used the stability of Earth-mass test particles to suggest that about half of
the known systems could contain a planet in the Habitable Zone.  As noted
above, the stability of a given region does not always imply that a planet
exists in the region -- the Asteroid Belt is a convenient example of a stable
region that contains no Earth-mass planet.  We suspect that the difference
between our estimate and that of \citet{jon05} are a number of systems in
which stable regions exist but planets are unlikely to form.  This is not
unexpected; for example, \citet{bar04} found stable regions for test particles
in four known extrasolar planetary systems, but \citet{ray06a} showed that
sizable terrestrial planets could only accrete in two of the systems.

Conversely, in previous work studying the formation of planets in systems with
one giant planet exterior to the terrestrial planets \citep{ray06} we
concluded that only 5\% of the known giant planet systems could form planets
of at least 0.3 $\mearth$ in the Habitable Zone.  Our current estimate has
increased to roughly one third, because of two factors: 1) our sample is
larger than in \citet{ray06}, and 2) we have determined that many inner giant
planet systems could form Earth-like planets in the Habitable Zone.  Given
that observational biases make it much easier to find close-in giant planets,
it is not surprising that our estimate for the number of habitable systems has
increased.

In this analysis, we have only considered a few key parameters -- stellar
mass, planet semi-major axis and eccentricity.  Table \ref{tab:xsp} also lists
the other parameters which we consider most important in terms of habitable
planet formation: planetary mass and stellar metallicity.  Perturbations from
lower (higher)-mass planets are correspondingly weaker (stronger), so for less
(more) massive giant planets our inner limit would move out (in), and our
outer limit would move in (out).  With the increasing number of lower-mass
extrasolar planets being discovered (e.g., \citet{but04,lov06}), we suspect that our
limits are relatively conservative for the population of extrasolar planets as a
whole.  The median mass of our sample is 0.76 $M_{Jup}$, so our limits are
appropriate on a statistical level.  The stellar metallicity is thought to be
related to the gas- to dust ratio of the disk and therefore the total mass in
terrestrial building blocks.  In addition, if the disk mass scales with
stellar mass (as is generally thought), there may be a deficit of rocky
material and perhaps fewer Earth-like planets around low-mass stars
\citep{ray06d}.

The next generations of space missions plan to discover and characterize
Earth-like planets around other stars.  These missions include NASA's {\it
Kepler}, {\it SIM} and {\it Terrestrial Planet Finder}, and CNES's {\it COROT}
and ESA's {\it Darwin}.  Our results suggest that terrestrial planets can coexist
with both close-in giant planets and giant planets in outer orbits, expanding the range of planetary systems that should be searched with these upcoming missions.  However, transit searches are very sensitive to orbital inclination.  In many cases our simulated Habitable Zone planets have mutual inclinations of 5-10$^{\circ}$ with respect to the orbit of the close-in giant planet, making detection of transits for both planets impossible.  Final inclinations are smaller for systems with an exterior giant planet, making systems with both an exterior and interior giant planet more amenable to detection, but overall the probability of seeing the transit of a close-in giant planet and a HZ planet in the same system is small.  However, transit timing measurements may be able to infer the presence
of HZ planets in these systems \citep{hol05,agol05}.

\section{Conclusion}
\label{sec:conc}
\subsection{Summary}
\label{sec:sum}
Our simulations demonstrate that terrestrial accretion can occur during and after giant planet migration on several fronts: 

\begin{enumerate}

\item Interior to the migrating giant planet, planetesimals and embryos are
shepherded inward by the combined effects of moving mean motion resonances and
dissipation via gas drag (\citet{tan99}; also shown by \citet{fog05} and
\citet{zho05}).  Rapid accretion occurs over the duration of migration (Stage
1), and $40\%-50\%$ of material inside 5 AU typically ends up in the form of
1 - 3 ``hot Earths'' when a gaseous disk is present.  These planets have masses
up to 5 $\mearth$ and accrete on the migration timescale of 10$^5$ years.  Hot
Earths tend to lie inside strong resonances such as the 2:1 and 3:2.
Planetesimals, which feel stronger drag, can be shepherded by higher-order
resonances.

\item Exterior to the giant planet, scattered embryos and planetesimals have
their orbits re-circularized by damping from gas drag (felt more strongly by
planetesimals) and dynamical friction (felt by embryos due to the
planetesimals) during Stage 2.  The embryos in this region begin with
compositions similar to the initial mean composition, but an influx of
material into the terrestrial zone (between the close-in giant and 2.5 AU)
from orbital decay of planetesimals due to gas drag can significantly enhance
the masses and fraction of water-rich material in these bodies.  From these
protoplanets a system of terrestrial planets forms on a $\sim10^{8}$ year
timescale after the gas disk has dissipated (Stage 3).

\item In the outer disk, damping via both gas drag and dynamical friction is
much weaker.  Thus, a scattered disk of planetesimals and embryos remain on
high eccentricity and inclination orbits, similar to the scattered disk of
comets in the solar system \citep{luu97, dun97}. Timescales for accretion are
very long.  The presence of a second giant planet serves to remove dynamically
hot bodies, diminishing the orbital excitation of bodies and reducing the
clearing time in the outer system.

\end{enumerate}

Planets formed in the Habitable Zone in a significant fraction of our
simulations.  These planets have masses and orbits similar to those seen in
previous simulations including only outer giant planets \citep{cha01,ray04,
ray06c}.  However, Habitable Zone planets in systems with close-in giant
planets tend to accrete a much larger amount of water than those in systems
with only outer giant planets \citep{ray06b}.  The reason for these high water
contents is twofold: 1) strong radial mixing is induced by the giant planet's
migration, and 2) in-spiraling icy planetesimals are easily accreted by
planets in the terrestrial zone.  Although we have not taken water depletion
into account, these planets contain about {\bf twenty} times as much water as
those formed in similar outer giant planet simulations \citep{ray04, ray06c}.
These planets are likely to be ``water worlds'', covered in kilometers-deep
global oceans \citep{leg04}.

The variations in the evolutionary process for each of the three models
examined here demonstrate the importance of both the configuration of giant
planets in the system as well as the presence and duration of the gaseous disk
in dictating the final parameters of the terrestrial planets formed.
Simulations run without including the gaseous disk (JSN) showed very little
planetesimal accretion during migration and almost no collisions after
migration due to the strong dynamical excitation of the disk (see Figure
\ref{fig:accret}).  Though embryo eccentricities and inclinations decreased
due to dynamical friction, the mean embryo composition remained similar to the
initial value and final terrestrial planet masses were in the lower end of the
mass range seen in the Solar System (see Figure \ref{fig:migall}).  In
contrast, embryos in simulations including the presence of a dissipating
gaseous disk (JD and JSD) accreted planetesimals and other embryos from both
the inner system ( during Stage 1) and the outer system (during Stage 2). This
resulted in both higher planet masses and water mass fractions, though this
effect was slightly mitigated in the JSD simulations due to more rapid
clearing in the outer system.  Eccentricities and inclinations are also
affected by the presence of an outer giant planet; in the JD simulations
dynamical excitation continues unabated during Stage 3, while in the JSD
simulations excited bodies are scattered by the Saturn-mass planet and the
remaining bodies therefore remain dynamically relaxed.

\subsection{Implications for Our Own System}
\label{sec:sum}
If planetary systems which suffer the migration of a giant planet to small distances can eventually form terrestrial planets similar to those in our own system, and the migration of young giant planets is a common result of interactions with the gaseous disk, then it is appropriate to consider the possibility that our own planetary system could have formed earlier generations of giant planets prior to those in the outer Solar System.  The systems simulated here undergo extensive radial mixing of large bodies throughout the inner and outer system through scattering, which could leave a elemental signature on the final population of both planetary and sub-planetary bodies.  However, radial mixing of dust and smaller bodies is also thought to occur through a variety of other processes in the planet formation region besides planetary scattering (see \citet{woo05} for a detailed discussion), and it is clear from the final abundances of planets in our simulations and others that any signature of the scattering of massive bodies early in the formation history will be quickly erased through accretion in the inner system due to continued mixing.  This is less true in the outer system, where long dynamical timescales and low solid and gas densities make accretion and radial transport less likely.  Therefore one of the primary constraints on giant planet formation and migration theories is the scattered population of Kuiper Belt objects (KBO's) and Oort Cloud comets -- both their orbital and compositional characteristics.  

Though conclusive detection of anomalous cometary and asteroidal abundances is naturally difficult and ambiguous, the detection of comet-like bodies with orbital characteristics suggesting an origin in the Oort Cloud but with physical appearances similar to asteroids could be considered a strong indicator of the scattering of massive inner-disk bodies by a migrating giant planet.  In fact, such bodies (known as Damocloids) have recently been discovered \citep{ash94,jew05}, but it is unclear whether any of these bodies are truly asteroidal in composition, and even if they are it has been shown that the stochastic nature of late-stage scattering by Jupiter could also produce small numbers of Oort Cloud comets of asteroidal nature \citep{weis97}.  A more conclusive sign of giant planet migration would be a classical KBO with a composition primarily composed of refractory materials; this would imply the re-circularization of a scattered inner-disk object, which would most likely only be possible in the presence of damping by significant amounts of gas or dust for long timescales.  In-depth analysis of the meteoritic and cometary record and other signatures of long-term planetary dynamics for evidence of early giant planet migration is clearly beyond the scope of this paper, and would most likely be inconclusive considering the many factors involved.  We must wait for the detection and accurate analysis of many more KBO's and Oort Cloud comets before we can place meaningful constraints on the possibility of Type II giant planet migration in our own system.

\subsection{Future Work}
\label{sec:fut}
In this study we have endeavored to utilize the most concrete data and plausible scenarios to develop our initial conditions for each model, while limiting the number of free parameters and ill-constrained variables.  However, there is naturally a range of parameter space in the formation of terrestrial planets in systems with a migrating giant planet that we could not explore.

We have only considered systems with one migrating giant planet.  The observed distribution of giant planet orbital parameters can be roughly reproduced by the combination of migration and gravitational scattering (e.g.,
\citet{ada03}.  Thus, in certain cases there may be multiple giant
planets migrating together.  Simulations by \citet{bry00} and \citet{kle00} suggest that if the planets attain their full mass while on closely-spaced orbits, the intervening gas between them would cause convergent migration, leading to resonance capture or scattering.  However, if one planet forms sufficiently early to migrate inwards, a second planet could form beyond the evacuation zone and migrate as well.  In this case gravitational scattering would be greatly enhanced
because inner disk material would interact with both giant planets and would
probably be scattered to larger orbital distances (as occurs for some bodies in model JSD).  However, in most cases the dynamic instabilities of the two massive planets would likely out-weigh the evolution of the smaller bodies, leading to chaotic scattering and removal of small bodies similar as outlined in \citet{ver05}.  If a combined migration leaves one planet in a small orbit within 0.5 AU and a second planet beyond 2 AU, one would expect a hybrid scenario between this work and \citet{ray06} with terrestrial planets forming in the stable region between the two giant planets, but more work is needed to confirm this.

Additional uncertainty surrounds the role of the gaseous disk.  In this work we bracket the potential range of disk masses with the JSD and JSN models.  As stated, the differences in the damping of scattering energies and the orbital decay of planetesimal orbits greatly impacts the final characteristics of the terrestrial planets that are formed.  To fully investigate the full range of gas disk properties and planetesimal size ranges would require an unreasonably large number of simulation runs. Our simulations were limited by computational resources.  We chose fixed values for several parameters such as the migration rate and disk properties, and chose to integrate these simulations for a long time rather than explore the effects of different parameters.  (Note that \citet{fog05} explored the effects of the disk's mass distribution.)  In addition, we did not include the effects of collisional fragmentation, which are certainly important for both planetesimal- (e.g., \citet{ben99} and embryo-scale collisions \citep{asp06}). More effects relating to the gas disk also remain to be modeled, including tidal gas drag (e.g., \citet{kom02}), which acts preferentially on larger bodies.  Clearly, more work remains to include all of the relevant physics.  Despite these limitations, our simulations are among the first to
realistically address this problem and examine the final results of terrestrial planet formation under these conditions (also see earlier work: \citet{man03,fog05,zho05}; Paper 1; \citet{fog06}).  Hopefully this work will encourage future studies to expand and improve the models explored here, both in our understanding of the relevant physical conditions and our ability to realistically model the myriad and complex forces at work in the formation of terrestrial planetary systems.

\section{Acknowledgments} 
\label{sec:ack}
We thank NASA Astrobiology Institute for funding, through the Penn State
(A.M. and S.S.), NASA Goddard (A.M.), Virtual Planetary Laboratory (S.R.) and
University of Colorado (S.R.) lead teams.  Thanks also to John Scalo for
providing our mass-luminosity relation.  S.N.R. was partially supported by an appointment to the NASA Postdoctoral Program at the University of Colorado Astrobiology Center, administered by Oak Ridge Associated Universities through a contract with NASA.

\clearpage

\clearpage

\clearpage

\clearpage

\clearpage

\clearpage

\clearpage

\clearpage

\clearpage

\clearpage

\clearpage

\clearpage
  
\clearpage
\begin{deluxetable}{lcccc} 
\tablewidth{0pt} 
\tablecaption{Embryo/Planetesimal Properties at the End of Stage 1} 
\tabletypesize{\scriptsize} 
\tablecolumns{5} 
\renewcommand{\arraystretch}{.6} 
\tablehead{
\colhead{} & \colhead{Zone\tablenotemark{1}} &
\colhead{JD} &  \colhead{JSD} & \colhead{JSN}}  
\startdata
  & Z.1 & 7.60$\mearth$ / 0.31$\mearth$ & 7.97$\mearth$ / 0.53$\mearth$ & 1.90$\mearth$ / 1.43$\mearth$\\
Tot. Mass: Em/Pl & Z.2 & 6.18$\mearth$ / 1.42$\mearth$ & 3.24$\mearth$ / 0.70$\mearth$ & 4.18$\mearth$ / 1.37$\mearth$\\
  & Z.3 & 2.77$\mearth$ / 0.07$\mearth$ & 4.93$\mearth$ / 0.50$\mearth$ & 3.36$\mearth$ / 1.79$\mearth$\\
\cline{1-5}\\
  & Z.1 & 0.45 (0.07 -- 4.35) & 0.46 (0.08 -- 4.90) & 0.18 (0.05 -- 0.54)\\
Em. M $(\mearth)$\tablenotemark{2} & Z.2 & 0.42 (0.08 -- 1.23) & 0.29 (0.09 -- 1.10) & 0.39 (0.05 -- 1.05)\\
  & Z.3 & 0.28 (0.09 -- 0.88) & 0.39 (0.10 -- 1.10) & 0.32 (0.03 -- 1.23)\\
\cline{1-5}\\
  & Z.1 & -2.09 (-5.00 -- -1.26) & -1.74 (-5.00 -- -0.30) & -1.75 (-5.00 -- -0.30)\\
Em. log(W.M.F.)\tablenotemark{2} & Z.2 & -0.56 (-5.00 -- -0.30) & -0.72 (-5.00 -- -0.30) & -0.59 (-5.00 -- -0.30)\\
  & Z.3 & -0.85 (-5.00 -- -0.30) & -0.60 (-5.00 -- -0.30) & -0.75 (-5.00 -- -0.30)\\
\cline{1-5}\\
  & Z.1 & -0.52 (-1.13 -- -0.37) & -0.54 (-1.18 -- -0.40) & -0.51 (-1.01 -- -0.34)\\
Em. log(Fe M.F.)\tablenotemark{2} & Z.2 & -0.76 (-1.23 -- -0.34) & -0.74 (-1.23 -- -0.37) & -0.74 (-1.23 -- -0.38)\\
  & Z.3 & -0.80 (-1.23 -- -0.49) & -0.81 (-1.23 -- -0.39) & -0.70 (-1.23 -- -0.37)\\
\cline{1-5}\\
  & Z.1 & 0.44 (0.01 -- 0.84) & 0.48 (0.01 -- 0.87) & 0.61 (0.02 -- 0.91)\\
Em. Eccen.\tablenotemark{2} & Z.2 & 0.37 (0.01 -- 0.92) & 0.42 (0.06 -- 0.86) & 0.44 (0.04 -- 0.94)\\
  & Z.3 & 0.75 (0.41 -- 0.97) & 0.62 (0.09 -- 0.99) & 0.71 (0.06 -- 0.98)\\
\cline{1-5}\\
  & Z.1 & 6.62 (0.22 -- 36.65) & 8.49 (0.20 -- 29.28) & 26.34 (0.51 -- 80.41)\\
Em. Inclin.$(^{\circ})$\tablenotemark{2} & Z.2 & 5.25 (0.44 -- 28.00) & 12.89 (0.90 -- 51.13) & 19.79 (1.26 -- 70.18)\\
  & Z.3 & 4.69 (0.64 -- 13.56) & 11.79 (1.23 -- 34.50) & 17.09 (1.40 -- 46.70)\\
\cline{1-5}\\
  & Z.1 & 0.00 (0.00 -- 0.10) & 0.01 (0.00 -- 0.19) & 0.54 (0.06 -- 0.94)\\
Pl. Eccen.\tablenotemark{2} & Z.2 & 0.09 (0.00 -- 0.23) & 0.08 (0.01 -- 0.40) & 0.41 (0.01 -- 0.97)\\
  & Z.3 & 0.11 (0.01 -- 0.21) & 0.31 (0.02 -- 0.79) & 0.60 (0.02 -- 0.99)\\
\cline{1-5}\\
  & Z.1 & 0.08 (0.01 -- 1.86) & 0.50 (0.00 -- 5.51) & 28.65 (0.19 -- 84.56)\\
Pl. Inclin.$(^{\circ})$\tablenotemark{2} & Z.2 & 2.58 (0.02 -- 10.52) & 3.67 (0.23 -- 18.00) & 17.93 (0.33 -- 80.30)\\
  & Z.3 & 2.36 (0.61 -- 5.00) & 5.69 (0.35 -- 32.37) & 14.94 (0.30 -- 81.01)\\
\enddata
\tablenotetext{1}{Radial zones are delineated as follows: Z.1 (a$<$5.2 AU), Z.2 (5.2$\le$a$<$9.5), and
	Z.3 (9.5$\le$a)} 
\tablenotetext{2}{Values represent the mean of the bodies in each zone, with the range of values
	in parentheses.}
\label{tab:st1}
\end{deluxetable}

\begin{deluxetable}{lcccc} 
\tablewidth{0pt} 
\tablecaption{Embryo/Planetesimal Properties at the End of Stage 2} 
\tabletypesize{\scriptsize} 
\tablecolumns{5} 
\renewcommand{\arraystretch}{.6} 
\tablehead{
\colhead{} & \colhead{Zone\tablenotemark{1}} &
\colhead{JD} &  \colhead{JSD} & \colhead{JSN}}  
\startdata
  & Z.1 & 9.25$\mearth$ / 1.01$\mearth$ & 5.77$\mearth$ / 0.51$\mearth$ & 1.34$\mearth$ / 0.82$\mearth$\\
Tot. Mass: Em/Pl & Z.2 & 5.12$\mearth$ / 0.02$\mearth$ & 0.63$\mearth$ / 0.01$\mearth$ & 0.67$\mearth$ / 0.19$\mearth$\\
  & Z.3 & 2.65$\mearth$ / 0.00$\mearth$ & 0.26$\mearth$ / 0.44$\mearth$ & 1.04$\mearth$ / 0.24$\mearth$\\
\cline{1-5}\\
  & Z.1 & 0.55 (0.09 -- 4.59) & 0.40 (0.08 -- 4.44) & 0.24 (0.11 -- 0.81)\\
Em. M $(\mearth)$\tablenotemark{2} & Z.2 & 0.46 (0.12 -- 1.59) & 0.42 (0.14 -- 0.95) & 0.38 (0.14 -- 0.60)\\
  & Z.3 & 0.32 (0.11 -- 0.93) & 0.52 (0.15 -- 0.90) & 0.52 (0.12 -- 1.05)\\
\cline{1-5}\\
  & Z.1 & -1.03 (-5.00 -- -0.30) & -1.03 (-5.00 -- -0.30) & -1.26 (-5.00 -- -0.30)\\
Em. log(W.M.F.)\tablenotemark{2} & Z.2 & -0.54 (-4.41 -- -0.30) & -0.57 (-2.74 -- -0.30) & -0.53 (-5.00 -- -0.30)\\
  & Z.3 & -0.72 (-5.00 -- -0.30) & -0.56 (-1.30 -- -0.30) & -0.50 (-5.00 -- -0.30)\\
\cline{1-5}\\
  & Z.1 & -0.59 (-1.13 -- -0.42) & -0.61 (-1.20 -- -0.40) & -0.56 (-1.01 -- -0.41)\\
Em. log(Fe M.F.)\tablenotemark{2} & Z.2 & -0.81 (-1.23 -- -0.48) & -0.83 (-1.13 -- -0.55) & -0.75 (-0.96 -- -0.48)\\
  & Z.3 & -0.86 (-1.23 -- -0.49) & -0.99 (-1.18 -- -0.86) & -0.79 (-1.22 -- -0.49)\\
\cline{1-5}\\
  & Z.1 & 0.13 (0.00 -- 0.55) & 0.07 (0.00 -- 0.64) & 0.34 (0.01 -- 0.79)\\
Em. Eccen.\tablenotemark{2} & Z.2 & 0.29 (0.02 -- 0.86) & 0.17 (0.01 -- 0.40) & 0.40 (0.11 -- 0.85)\\
  & Z.3 & 0.53 (0.17 -- 0.93) & 0.73 (0.73 -- 0.74) & 0.42 (0.15 -- 0.80)\\
\cline{1-5}\\
  & Z.1 & 2.63 (0.03 -- 16.09) & 2.07 (0.05 -- 26.80) & 20.87 (3.00 -- 48.33)\\
Em. Inclin.$(^{\circ})$\tablenotemark{2} & Z.2 & 6.85 (0.72 -- 21.31) & 5.53 (0.91 -- 13.54) & 32.45 (7.42 -- 55.09)\\
  & Z.3 & 6.77 (0.83 -- 16.80) & 21.40 (15.11 -- 30.44) & 22.97 (5.55 -- 60.20)\\
\cline{1-5}\\
  & Z.1 & 0.06 (0.00 -- 0.31) & 0.03 (0.00 -- 0.31) & 0.47 (0.04 -- 0.85)\\
Pl. Eccen.\tablenotemark{2} & Z.2 & 0.16 (0.05 -- 0.25) & 0.05 (0.02 -- 0.11) & 0.41 (0.03 -- 0.92)\\
  & Z.3 & 0.21 (0.18 -- 0.23) & 0.05 (0.00 -- 0.29) & 0.57 (0.06 -- 0.97)\\
\cline{1-5}\\
  & Z.1 & 1.48 (0.01 -- 12.63) & 1.18 (0.05 -- 6.29) & 27.68 (0.74 -- 69.72)\\
Pl. Inclin.$(^{\circ})$\tablenotemark{2} & Z.2 & 8.03 (1.69 -- 11.38) & 1.73 (1.41 -- 2.56) & 27.78 (3.14 -- 66.70)\\
  & Z.3 & 6.54 (4.21 -- 9.10) & 1.82 (0.24 -- 18.63) & 27.09 (1.04 -- 71.22)\\
\enddata
\tablenotetext{1}{Radial zones are delineated as follows: Z.1 (a$<$5.2 AU), Z.2 (5.2$\le$a$<$9.5), and
	Z.3 (9.5$\le$a)} 
\tablenotetext{2}{Values represent the mean of the bodies in each zone, with the range of values
	in parentheses.}
\label{tab:st2}
\end{deluxetable}

\begin{deluxetable}{lcccc} 
\tablewidth{0pt} 
\tablecaption{Final Embryo/Planetesimal Properties} 
\tabletypesize{\scriptsize} 
\tablecolumns{5} 
\renewcommand{\arraystretch}{.6} 
\tablehead{
\colhead{} & \colhead{Zone\tablenotemark{1}} &
\colhead{JD} &  \colhead{JSD} & \colhead{JSN}}  
\startdata
  & Z.1 & 5.52$\mearth$ / 0.05$\mearth$ & 4.08$\mearth$ / 0.05$\mearth$ & 1.17$\mearth$ / 0.19$\mearth$\\
Tot. Mass: Em/Pl & Z.2 & 2.71$\mearth$ / 0.00$\mearth$ & 0.26$\mearth$ / 0.00$\mearth$ & 0.00$\mearth$ / 0.01$\mearth$\\
  & Z.3 & 4.83$\mearth$ / 0.02$\mearth$ & 0.00$\mearth$ / 0.35$\mearth$ & 0.51$\mearth$ / 0.04$\mearth$\\
\cline{1-5}\\
  & Z.1 & 1.03 (0.12 -- 4.15) & 0.58 (0.08 -- 1.77) & 0.29 (0.11 -- 0.82)\\
Em. M $(\mearth)$\tablenotemark{2} & Z.2 & 0.48 (0.14 -- 1.59) & 0.35 (0.17 -- 0.46) & 0.00 (0.00 -- 0.00)\\
  & Z.3 & 0.32 (0.09 -- 1.24) & 0.00 (0.00 -- 0.00) & 1.03 (1.01 -- 1.05)\\
\cline{1-5}\\
  & Z.1 & -0.68 (-2.93 -- -0.30) & -0.87 (-5.00 -- -0.30) & -1.01 (-5.00 -- -0.30)\\
Em. log(W.M.F.)\tablenotemark{2} & Z.2 & -0.56 (-5.00 -- -0.30) & -0.71 (-1.46 -- -0.30) & 0.00 (0.00 -- 0.00)\\
  & Z.3 & -0.67 (-5.00 -- -0.30) & 0.00 (0.00 -- 0.00) & -0.30 (-0.30 -- -0.30)\\
\cline{1-5}\\
  & Z.1 & -0.65 (-0.96 -- -0.50) & -0.65 (-1.19 -- -0.44) & -0.55 (-0.94 -- -0.42)\\
Em. log(Fe M.F.)\tablenotemark{2} & Z.2 & -0.81 (-1.21 -- -0.54) & -0.69 (-0.88 -- -0.59) & 0.00 (0.00 -- 0.00)\\
  & Z.3 & -0.81 (-1.23 -- -0.48) & 0.00 (0.00 -- 0.00) & -0.91 (-0.93 -- -0.90)\\
\cline{1-5}\\
  & Z.1 & 0.29 (0.01 -- 0.74) & 0.17 (0.00 -- 0.57) & 0.24 (0.03 -- 0.42)\\
Em. Eccen.\tablenotemark{2} & Z.2 & 0.30 (0.08 -- 0.58) & 0.15 (0.12 -- 0.35) & 0.00 (0.00 -- 0.00)\\
  & Z.3 & 0.39 (0.06 -- 0.95) & 0.00 (0.00 -- 0.00) & 0.19 (0.13 -- 0.24)\\
\cline{1-5}\\
  & Z.1 & 14.40 (0.69 -- 32.24) & 6.66 (0.35 -- 21.80) & 17.27 (3.71 -- 28.34)\\
Em. Inclin.$(^{\circ})$\tablenotemark{2} & Z.2 & 18.46 (4.73 -- 42.86) & 13.66 (6.50 -- 19.22) & 0.00 (0.00 -- 0.00)\\
  & Z.3 & 19.76 (2.71 -- 50.25) & 0.00 (0.00 -- 0.00) & 3.21 (2.74 -- 3.82)\\
\cline{1-5}\\
  & Z.1 & 0.13 (0.05 -- 0.34) & 0.44 (0.03 -- 0.89) & 0.40 (0.06 -- 0.81)\\
Pl. Eccen.\tablenotemark{2} & Z.2 & 0.67 (0.67 -- 0.67) & 0.37 (0.37 -- 0.37) & 0.43 (0.17 -- 0.73)\\
  & Z.3 & 0.69 (0.40 -- 0.90) & 0.08 (0.01 -- 0.26) & 0.28 (0.06 -- 0.60)\\
\cline{1-5}\\
  & Z.1 & 8.80 (2.28 -- 50.79) & 31.25 (2.02 -- 61.23) & 31.65 (2.02 -- 59.38)\\
Pl. Inclin.$(^{\circ})$\tablenotemark{2} & Z.2 & 93.39 (93.39 -- 93.39) & 43.84 (43.84 -- 43.84) & 31.21 (8.42 -- 48.70)\\
  & Z.3 & 33.94 (12.39 -- 92.93) & 4.07 (0.21 -- 27.00) & 18.87 (4.03 -- 42.56)\\
\enddata
\tablenotetext{1}{Radial zones are delineated as follows: Z.1 (a$<$5.2 AU), Z.2 (5.2$\le$a$<$9.5), and
	Z.3 (9.5$\le$a)} 
\tablenotetext{2}{Values represent the mean of the bodies in each zone, with the range of values
	in parentheses.}
\label{tab:st3}
\end{deluxetable}

\clearpage
\scriptsize
\begin{deluxetable}{lccccccc}
\tablewidth{0pt}
\tablecaption{Properties of planets formed in JD Simulations\tablenotemark{1}}
%\tabletypesize{\scriptsize}
\tablecolumns{9}
\renewcommand{\arraystretch}{.6}
\tablehead{
\colhead{Simulation} &  
\colhead{$a (AU)$} &
\colhead{$\bar{e}$\tablenotemark{2}} &
\colhead{$\bar{i} (^{\circ})$} &
\colhead{M$(\mearth)$} &
\colhead{W.M.F.} &
\colhead{Fe M.F.}}
\startdata
 JD-1 & \bf 0.10 & \bf  0.01 &  \bf 0.6  & \bf  323.42  & \bf --  & \bf -- \\
 &     0.64 &  0.33 &  1.5  &    1.14  &  $7.72\times 10^{-2}$  &    0.27\\
 &     1.80 &  0.33 &  5.5  &    2.14  &  $1.13\times 10^{-1}$ &     0.27\\
 &     3.95 &  0.47 &  7.2  &    0.90  & $5.00\times 10^{-1}$ &     0.11\\
 &     4.01 &  0.46 &  24.4 &     0.24 &  $1.00\times10^{-5}$    &     0.29\\
\\ JD-3 &   \bf     0.14  &   \bf  0.01  &  \bf    0.5  &  \bf 326.45 &  \bf --  & \bf -- \\
 &     1.74  &    0.46  &    38.0  &    0.32 & $5.78\times 10^{-2}$  &    0.29\\
 &     2.27  &    0.48  &    22.9  &    0.36 & $1.01\times 10^{-1}$  &    0.27\\
 &     3.80  &    0.27  &     9.0  &    1.25 & $4.05\times 10^{-1}$  &    0.16\\
\\ JD-4 &   \bf    0.14  &  \bf   0.02   &   \bf  0.3 &  \bf  324.17 &  \bf --  &  \bf --\\
 &      0.31  &    0.06   &    1.4 &     1.63 & $4.60\times 10^{-2}$   &   0.31\\
 &      1.19  &    0.12   &    8.0 &     0.92 & $7.64\times 10^{-2}$   &   0.27\\
 &      2.47  &    0.59   &   42.7 &     0.30 & $1.52\times 10^{-1}$   &   0.24\\
 &      2.86  &    0.21   &   10.5 &     1.06 & $7.08\times 10^{-2}$   &   0.20\\
 &      3.33  &    0.31   &   24.0 &     0.23 & $9.23\times 10^{-2}$   &   0.28\\
 &      3.40  &    0.19   &   12.5 &     0.83 & $5.00\times 10^{-1}$   &   0.11\\
 &      3.67  &    0.21   &   16.3 &     0.14 & $7.39\times 10^{-2}$   &   0.26\\
 &      3.80  &    0.69   &   45.7 &     0.14 & $2.28\times 10^{-1}$   &   0.22\\
 &      3.92  &    0.23   &   11.2 &     0.39 & $9.03\times 10^{-2}$   &   0.26\\
 &      4.60  &    0.32   &   10.3 &     0.46 & $1.79\times 10^{-1}$   &   0.17\\
\\ JD-5 &     0.12   &   0.01  &     0.7  &    4.15 & $2.34\times 10^{-2}$   &   0.25\\
 &     \bf  0.21   &  \bf  0.01  &    \bf  0.1  &  \bf 320.92 & \bf --   & \bf -- \\
 &      0.91   &   0.08  &     5.0  &    3.05 & $9.23\times 10^{-2}$   &   0.27\\
 &      2.98   &   0.33  &    13.3  &    0.55 & $1.36\times 10^{-1}$   &   0.26\\
 &      3.38   &   0.22  &    11.9  &    0.32 & $1.15\times 10^{-3}$   &   0.31\\
 &      4.32   &   0.51  &    27.7  &    0.12 & $3.77\times 10^{-3}$   &   0.27\\
 &      4.78   &   0.27  &    14.9  &    0.84 & $5.00\times 10^{-1}$   &   0.13\\
 &      4.68   &   0.14  &    13.5  &    0.72 & $5.00\times 10^{-1}$   &   0.11\\
\enddata
\tablenotetext{1}{Planets are defined to be $> 0.1 \mearth$ and inside 5 AU.
  Close-in giant planets are shown in bold.}
\tablenotetext{2}{Orbital elements are averaged over the last Myr of
 the simulation.}
 \label{tab:jd}
\end{deluxetable}
\normalsize

\scriptsize
\begin{deluxetable}{lccccccc}
\tablewidth{0pt}
\tablecaption{Properties of planets formed in JSD Simulations}
\tablecolumns{9}
\renewcommand{\arraystretch}{.6}
\tablehead{
\colhead{Simulation} &  
\colhead{$a (AU)$} &
\colhead{$\bar{e}$} &
\colhead{$\bar{i} (^{\circ})$} &
\colhead{M$(\mearth)$} &
\colhead{W.M.F.} &
\colhead{Fe M.F.}}
\startdata
JSD-1 &   \bf 0.08   & \bf  0.00   & \bf   0.2  & \bf 325.57 & \bf --   & \bf  -- \\
 &        0.19   &   0.01   &    0.5  &    0.46 & $1.12\times 10^{-1}$   &   0.25\\
 &        0.86   &   0.12   &    4.1  &    0.46 & $4.76\times 10^{-2}$   &   0.20\\
 &        1.62   &   0.07   &    2.0  &    1.46 & $4.34\times 10^{-2}$   &   0.23\\
 &        2.32   &   0.22   &    3.4  &    0.46 & $2.59\times 10^{-2}$   &   0.30\\
 &        3.38   &   0.08   &    2.6  &    0.29 & $7.68\times 10^{-2}$   &   0.23\\
 &        4.57   &   0.12   &    4.6  &    0.46 & $5.00\times 10^{-1}$   &   0.13\\
\\ JSD-2 &  \bf   0.06  &  \bf  0.00  &  \bf   2.3  & \bf 325.67 & \bf -- & \bf --\\
  &       0.14  &    0.02  &     2.3  &    0.38 & $9.99\times 10^{-2}$   &   0.13\\
  &       0.26  &    0.00  &     2.3  &    0.14 & $5.55\times 10^{-2}$   &   0.28\\
  &       0.75  &    0.29  &    17.2  &    0.41 & $1.43\times 10^{-1}$   &   0.23\\
  &       1.42  &    0.18  &     7.1  &    1.77 & $1.24\times 10^{-1}$   &   0.27\\
  &       3.09  &    0.23  &     5.1  &    0.95 & $5.00\times 10^{-1}$   &   0.12\\
  &       4.06  &    0.32  &     8.9  &    0.60 & $5.59\times 10^{-2}$   &   0.23\\
\\ JSD-3 &  0.07   &   0.01   &    0.7  &    1.24 & $3.25\times 10^{-2}$  &    0.25\\
  &       0.11   &   0.00   &    0.6  &    3.23 & $1.92\times 10^{-2}$  &    0.31\\
  &      \bf  0.19   &  \bf  0.00   &   \bf  0.6  &  \bf 320.21 & \bf --  &  \bf -- \\
  &       0.41   &   0.03   &    1.4  &    0.12 & $1.00\times10^{-5}$   &      0.35\\
  &       0.51   &   0.01   &    1.0  &    0.68 & $3.67\times 10^{-2}$  &    0.32\\
  &       0.61   &   0.01   &    1.1  &    0.74 & $4.14\times 10^{-2}$  &    0.32\\
  &       1.17   &   0.02   &    0.8  &    0.24 & $7.57\times 10^{-2}$  &    0.23\\
   &      2.34   &   0.12   &    4.1  &    0.19 & $3.70\times 10^{-2}$  &    0.28\\
   &      2.63   &   0.17   &    6.5  &    0.13 & $2.90\times 10^{-2}$  &    0.31\\
   &      3.67   &   0.10   &    4.8  &    0.33 & $6.68\times 10^{-2}$  &    0.16\\
   &      3.93   &   0.18   &   15.0  &    0.17 & $6.81\times 10^{-2}$  &    0.20\\
   &      4.49   &   0.17   &    2.7  &    0.61 & $5.00\times 10^{-1}$  &    0.12\\
\\ JSD-4 &   \bf  0.05   &  \bf  0.00   &   \bf  2.1 &  \bf  324.82 & \bf -- & \bf -- \\
 &   0.25   &   0.01   &    2.1 &     1.11 & $5.20\times 10^{-2}$   &   0.20\\
 &   0.55   &   0.14   &    8.1 &     0.77 & $1.05\times 10^{-1}$   &   0.19\\
 &   0.92   &   0.14   &    6.8 &     0.36 & $1.37\times 10^{-1}$   &   0.27\\
 &   1.26   &   0.07   &    7.1 &     0.63 & $1.16\times 10^{-1}$   &   0.21\\
 &   2.65   &   0.20   &   19.8 &     0.11 & $3.33\times 10^{-3}$   &   0.27\\
 &   3.02   &   0.04   &    2.0 &     1.32 & $3.25\times 10^{-1}$   &   0.17\\
\enddata
\label{tab:jsd}
\end{deluxetable}
\normalsize

\scriptsize
\begin{deluxetable}{lccccccc}
\tablewidth{0pt}
\tablecaption{Properties of planets formed in JSN Simulations}
\tablecolumns{9}
\renewcommand{\arraystretch}{.6}
\tablehead{
\colhead{Simulation} &  
\colhead{$a (AU)$} &
\colhead{$\bar{e}$} &
\colhead{$\bar{i} (^{\circ})$} &
\colhead{M$(\mearth)$} &
\colhead{W.M.F.} &
\colhead{Fe M.F.}}
\startdata
JSN-2 &  \bf   0.24  &  \bf  0.07  &  \bf   1.1 &  \bf 318.85  & --- & --- \\
&    1.13  &    0.35  &    24.9 &     0.19 & 1.84$\times 10^{-3}$ &  0.30 \\
&    1.32  &    0.20  &    17.0 &     0.28 & 1.26$\times 10^{-3}$ &  0.35 \\
&    3.98  &    0.06  &    10.2 &     0.53 & 4.93$\times 10^{-1}$ &  0.13 \\
\\ JSN-3 &   0.05  &  0.37  &   29.3 &   0.22 & 4.99$\times 10^{-1}$ &   0.11 \\
 &    \bf  0.24  & \bf   0.05  &  \bf  3.7 &  \bf  318.82 & --- & --- \\
 &    1.70  &    0.25  &     3.8 &     0.24 & 4.85$\times 10^{-2}$ &   0.24 \\
 &    2.26  &    0.04  &     7.0 &     0.18 & 1$\times 10^{-5}$    &   0.30 \\
 &    4.04  &    0.40  &    21.2 &     0.22 & 1$\times 10^{-5}$    &   0.29 \\
 &    4.77  &    0.18  &    14.0 &     0.54 & 5.00$\times 10^{-1}$ &   0.14 \\
 &    22.70  &    0.25  &     3.8 &     1.01 & 5.00$\times 10^{-1}$ &   0.12 \\     
\\ JSN-4 & \bf  0.24  &  \bf  0.04  &  \bf   1.7 &  \bf 319.29  & --- & --- \\
 &   0.96  &    0.07  &    17.3 &     0.22 & 2.46$\times 10^{-3}$ &  0.28 \\
 &   2.76  &    0.04  &    28.9 &     0.30 & 2.13$\times 10^{-3}$ &  0.26 \\
 &   4.50  &    0.22  &    12.4 &     0.82 & 5.00$\times 10^{-1}$ &  0.11 \\
 &   18.53  &    0.13  &     3.6 &     1.05 & 5.00$\times 10^{-1}$ &  0.13 \\
\\ JSN-5 & \bf   0.24  &  \bf  0.04  & \bf    2.0 & \bf  319.10  & --- & --- \\
 &   1.10  &    0.21  &    12.0 &     0.40 & 1$\times 10^{-5} $   &  0.36 \\
 &   1.15  &    0.31  &    24.6 &     0.13 & 5.44$\times 10^{-3}$ &  0.29 \\
 &   2.68  &    0.28  &    13.3 &     0.17 & 2.02$\times 10^{-3}$ &  0.38 \\
 &   3.39  &    0.47  &    23.4 &     0.11 & 1$\times 10^{-5} $   &  0.31 \\
 &   3.59  &    0.24  &    20.0 &     0.12 & 1$\times 10^{-5} $   &  0.32 \\
\enddata
\label{tab:jsn}
\end{deluxetable}
\normalsize

\clearpage
\begin{deluxetable}{ccccc} 
\tablewidth{0pt} 
\tablecaption{Giant Planet Semi-Major Axis Limits for Potentially Habitable Systems} 
\tabletypesize{\scriptsize} 
\tablecolumns{5} 
\renewcommand{\arraystretch}{.6} 
\tablehead{
\colhead{M$_\star$ (M$_\odot$)} &  
\colhead{Sp. Type\tablenotemark{1}} & 
\colhead{Hab Zone (AU)\tablenotemark{2}} &  
\colhead{Inner Limit (AU)} & 
\colhead{Outer Limit (AU)}} 
\startdata
0.1 & M6 & 0.024 - 0.045 & 0.015 & 0.075\\
0.4 & M3 & 0.10 - 0.19 & 0.06 & 0.32\\
0.7 & K6 & 0.28 - 0.52 & 0.17 & 0.87\\
1.0 & G2 & 0.8 - 1.5 & 0.5 & 2.5 \\
1.3 & F8 & 2.3 - 4.3 & 1.45 & 7.2 \\
1.6 & F0 & 6.5 - 12.3 & 4.1 & 20.5\\
2.0 & A5 & 25 - 47 & 15.7 & 78.3\\
\enddata
 \tablenotetext{1}{Spectral types from Table 8.1 of \cite{rei00} and Appendix
E from \cite{car96}. Spectral types of low-mass stars are age-dependent.}
\tablenotetext{2}{Habitable Zones scaled by $L_{\star}^{1/2}$to 0.8 - 1.5 AU for a solar-mass
star.}
\label{tab:gplim}
\end{deluxetable}
   
\clearpage
\begin{deluxetable}{lcccccc}
 \tablewidth{0pt}
 \tablecaption{Potentially Habitable Exoplanet Systems\tablenotemark{1}}
 \tabletypesize{\scriptsize}
 \tablecolumns{7}
 \renewcommand{\arraystretch}{.6}
 \tablehead{
 \colhead{System} &  
 \colhead{M$_\star$ (M$_\odot$)} & 
 \colhead{[Fe/H]} & 
 \colhead{M$_{pl}$ (M$_{J}$)} & 
 \colhead{$a$ (AU)} &
 \colhead{$e$}  &
 \colhead{HZ}}
 \startdata
 OGLE-05-071L &  0.13 & --- &  0.9 &    1.800 &     --- &     0.03-0.06  \\
 OGLE-05-390L &  0.22 &  --- &  0.02 &    2.100 &     --- &     0.06-0.10 \\
      GJ 581 &  0.31 & -0.25 &  0.052 &    0.041 &     0.00 &     0.08-0.14  \\
 OGLE235-MOA53 &  0.36 &  --- &  1.5 &    3.000 &     --- &     0.09-0.17   \\
  HD 41004 A &  0.40 & 0.16 &  17.892 &    0.018 &     0.08 &     0.10-0.19 \\
 OGLE-05-169L &  0.49 & --- &  0.04 &    2.800 &     --- &     0.14-0.25 \\
   HD 330075 &  0.70 & 0.08 &  0.623 &    0.039 &     0.00 &     0.28-0.52 \\
    HD 27894 &  0.75 & 0.3 &  0.618 &    0.122 &     0.05 &     0.33-0.62  \\
   HD 114386 &  0.76 & 0.004 & 1.343 &    1.714 &     0.23 &     0.34-0.64  \\
    HD 13445 &  0.77 & -0.27 &  3.9 &    0.113 &     0.04 &     0.35-0.66  \\
  OGLE-TR-113 &  0.77 &  0.14 &  1.35 &    0.023 &     0.00 &     0.35-0.66  \\
   HD 111232 &  0.78 & -0.36 &  6.803 &    1.975 &     0.20 &     0.37-0.69  \\
    HD 63454 &  0.80 & 0.11 &  0.385 &    0.036 &     0.00 &     0.39-0.74  \\
   HD 192263 &  0.81 & 0.05 &  0.641 &    0.153 &     0.05 &     0.41-0.76  \\
  OGLE-TR-111 &  0.81 & 0.12 &  0.52 &    0.047 &     0.00 &     0.41-0.76  \\
 Eps Eri &  0.82 & -0.03 &  1.058 &    3.377 &     0.25 &     0.42-0.79  \\
   HD 189733 &  0.82 & -0.03 &  1.152 &    0.031 &     0.00 &     0.42-0.79  \\
   HD 130322 &  0.88 & 0.006 &  1.088 &    0.091 &     0.02 &     0.52-0.98  \\
     TrES-1 &  0.89 & 0.001 &  0.759 &    0.039 &     0.00 &     0.54-1.01  \\
     HD 4308 &  0.90 & -0.31 &  0.047 &    0.118 &     0.00 &     0.56-1.05  \\
     55 Cnc e &  0.91 &  0.31 &  0.038 &    0.038 &     0.09 &     0.58-1.09  \\
     55 Cnc b &   & &  0.833 &    0.114 &     0.01 &     0.58-1.09  \\
     55 Cnc c &   & &  0.157 &    0.238 &     0.07 &     0.58-1.09  \\
     55 Cnc d &   & &  3.887 &    5.964 &     0.09 &     0.58-1.09  \\
    HD 46375 &  0.92 & 0.24 &  0.226 &    0.040 &     0.06 &     0.60-1.13  \\
     HD 2638 &  0.93 & 0.16 &  0.477 &    0.044 &     0.00 &     0.62-1.17  \\
   HD 102195 &  0.93 & -0.09 &  0.492 &    0.049 &     0.06 &     0.62-1.17  \\
   HD 164922 &  0.94 & 0.17 &  0.36 &    2.110 &     0.05 &     0.65-1.21  \\
    HD 89307 &  1.00 & -0.16 & 2.601 &    3.945 &     0.01 &     0.80-1.50  \\
    HD 83443 &  1.00 & 0.36 &  0.398 &    0.041 &     0.01 &     0.80-1.50  \\
    Rho Cnc B &  1.00 & -0.20 &  1.092 &    0.229 &     0.06 &     0.80-1.50  \\
  BD -10 3166 &  1.01 & 0.38 &  0.458 &    0.045 &     0.02 &     0.83-1.55  \\
    HD 70642 &  1.05 & 0.16 &  1.97 &    3.230 &     0.03 &     0.96-1.79  \\
    HD 73256 &  1.05 & 0.26 &  1.867 &    0.037 &     0.03 &     0.96-1.79  \\
   HD 212301 &  1.05 & -0.18 & 0.396 &    0.034 &     0.00 &     0.96-1.79  \\
    HD 49674 &  1.06 & 0.31 &  0.105 &    0.058 &     0.09 &     0.99-1.86  \\
   HD 195019 &  1.07 & 0.07 &  3.681 &    0.139 &     0.01 &     1.03-1.92  \\
   HD 187123 &  1.08 & 0.12 &  0.527 &    0.043 &     0.02 &     1.06-1.99  \\
     51 Peg &  1.09 & 0.20 &  0.472 &    0.053 &     0.01 &     1.10-2.07  \\
   HD 107148 &  1.12 & 0.31 &  0.21 &    0.269 &     0.05 &     1.23-2.30  \\
     HD 76700 &  1.13 & 0.35 &  0.232 &    0.051 &     0.09 &     1.27-2.38  \\
   HD 209458 &  1.14 & 0.014 &  0.689 &    0.047 &     0.00 &     1.31-2.47  \\
   OGLE-TR-10 &  1.17 & 0.12 &  0.63 &    0.042 &     0.00 &     1.46-2.74  \\
   OGLE-TR-56 &  1.17 & --- &  1.24 &    0.023 &     --- &     1.46-2.74  \\
   HD 121504 &  1.18 & 0.16 &  1.221 &    0.329 &     0.03 &     1.51-2.84  \\
   HD 149143 &  1.20 & 0.25 &  1.327 &    0.053 &     0.00 &     1.63-3.05  \\
    HD 75289 &  1.21 & 0.22 &  0.466 &    0.048 &     0.03 &     1.68-3.16  \\
   HD 109749 &  1.21 & 0.26 &  0.277 &    0.063 &     0.00 &     1.68-3.16  \\
   HD 179949 &  1.21 & 0.14 &  0.916 &    0.044 &     0.02 &     1.68-3.16  \\
    HD 86081 &  1.21 & 0.26 &  1.49 &    0.035 &     0.01 &     1.68-3.16  \\
   HD 149026 &  1.30 & 0.36 &  0.337 &    0.043 &     0.00 &     2.31-4.34  \\
   HD 224693 &  1.33 & 0.34 &  0.718 &    0.192 &     0.04 &     2.57-4.82  \\
    Tau Boo &  1.35 & 0.23 &  4.126 &    0.048 &     0.02 &     2.76-5.17  \\
  OGLE-TR-132 &  1.35 & 0.43 &  1.19 &    0.031 &     0.00 &     2.76-5.17  \\
   HD 177830 &  1.46 & 0.54 &  1.531 &    1.227 &     0.10 &     4.04-7.58  \\
 Eps Ret &  1.49 & 0.42 &  1.556 &    1.270 &     0.06 &     4.48-8.41  \\
   HD 104985 &  1.60 & -0.35 &  6.315 &    0.779 &     0.03 &     6.55-12.27  \\
\enddata
\tablenotetext{1}{Data from Butler et al (2006), www.exoplanet.eu, and  references therein.}
\label{tab:xsp}
\end{deluxetable}

\end{document}